\documentclass[12pt]{article}%
\usepackage{graphicx} 
\usepackage{amsmath}
\usepackage{mathtools}
\usepackage[onehalfspacing]{setspace}
\usepackage{multicol}
\usepackage{natbib}
\usepackage{amsfonts}
\usepackage{amssymb}
\usepackage{graphicx}
\usepackage{caption}
\usepackage{xcolor}
\usepackage{multirow}
\usepackage{array}
\usepackage[colorlinks=true,linkcolor=blue,citecolor=blue]{hyperref}
\usepackage{comment}
\usepackage{enumerate}
\usepackage{nicematrix}
\usepackage{booktabs}
\usepackage{lscape}
\usepackage[a4paper, total={6in, 8in}, margin=1.25in]{geometry}

\usepackage{float}%

\title{Heterogeneous Bribery, Technology Choice, and Capital Accumulation\thanks{We would like to thank Hung-Ju Chen, Yili Chien, William Dupor, Yiting Li, Gabriel Mihalache, Julian Neira, Ping Wang, C.C. Yang, and participants at the 18th Annual Dynare Conference, 3rd RISE Workshop, SAET 2025 Conference, Kobe University, National Taiwan University for helpful comments and suggestions. We would like to thank Farid Farrokhi for sharing the code. Declarations of interest: none.}}

\author {Jafar M. Olimov\thanks{Ohio State University. Email: olimov@gmail.com } \and Yi-Chan Tsai\thanks{%
National Taiwan University, Email: yichantsai@ntu.edu.tw }
\and Hao-Yu Yang\thanks{%
National Taiwan University, Email: r11323026@ntu.edu.tw }} 
\date{}
\date{\today }

\begin{document}
\maketitle

\begin{abstract}
We study the production, entry, and technological decisions of firms in the presence of bribery. We find that bribery can be justified even in the absence of bureaucratic inefficiencies. We document substantial technology-specific heterogeneity in bribery in 148 countries and incorporate it into a general equilibrium model, where firms use capital-intensive or labor-intensive technology. When bribery more heavily affects less efficient labor-intensive firms, resources move toward more efficient capital-intensive firms, resulting in higher capital accumulation and aggregate output. In poorer countries, the elimination of bribery only for capital-intensive firms increases the capital stock by 18.7\% more and the aggregate output by 3.4\% more than the complete elimination of bribery. In wealthier countries, the elimination of bribery only for capital-intensive firms increases the capital stock by 44.4\% more and the aggregate output by 15.4\% more than the complete elimination of bribery. Our findings challenge the established view of bribery as uniformly harmful and demonstrate how the within-country heterogeneity in bribery can explain cross-country differences in income.



\end{abstract}

\newpage

\section{Introduction}
Although there is a nearly universal consensus that corruption and bribery are harmful to economic performance, there are arguments to justify bribery in the second best world\footnote{See, for example, \cite{Bardhan1997}, \cite{Ehrlich1999}, \cite{MeonSekkat2005}, \cite{Uberti2022}.}\footnote{In the paper, we use the terms ``bribery" and ``corruption" interchangeably. ``Corruption" is a more general term than ``bribery" and includes other forms of bureaucratic failure such as red tape.}. For example, \cite{AcemogluVerdier2000} argue that some bureaucratic bribery is justified when a benevolent government needs to correct market failures that require intervention by self-interested bureaucrats whose actions cannot be fully monitored. The literature, originating with \cite{Huntington1968}, argues that bribery can increase overall efficiency by allowing more efficient firms to circumvent restrictive bureaucratic barriers or to gain access to scarce public resources through bribe payments (e.g. \cite{Lui1985}, \cite{MeonWeil2010}, \cite{Weaver2021}). In all these arguments, bribery is viewed as the necessary means to alleviate more costly economic problems. 

We argue that bribery can be justified even in the absence of bureaucratic inefficiencies. If bribery is to be viewed (i) as a source of resource misallocation for firms, and (ii) bribery rates are differentiated by types of firms or by different sectors of economy, then more productive sectors of economy with less bribery can attract more firms and resources away from less productive sectors of economy with more bribery. In this way, the presence of bribery, differentiated by sectors of the economy, can serve as a source of higher capital accumulation and ultimately higher aggregate output. 

To support this argument, we document substantial firm-level heterogeneity in bribery rates within and across individual countries. Specifically, we present empirical evidence that in low-income countries, the bribery rate of modern firms (operating with capital-intensive production technology) is 33\% higher than the bribery rate of traditional firms (operating with labor-intensive production technology)\footnote{Countries with GDP per capita in the lowest third of the sample are classified as low-income; those in the middle third as middle-income; and those in the top third as high-income.}. In high-income countries, the pattern of bribery is reversed: the bribery rate of modern firms is 40\% less than the bribery rate of traditional firms. In middle-income countries, the bribery rates of modern and traditional firms are similar: the bribery rate of modern firms is 10.8\% more than that of traditional firms (see Table \ref{tab:summary_stats})\footnote{The differences in bribery rates between modern and traditional firms are calculated as percentage differences between average bribes (as a share of total sales) of all modern firms and all traditional firms that report paying positive bribes.}. We argue that this previously overlooked within-country heterogeneity in bribery rates can potentially explain large cross-country differences in capital accumulation, technology adoption, and aggregate output. 

We incorporate the firm-level heterogeneity in bribery rates in a dynamic stochastic general equilibrium model, where firms endogenously make entry decisions, face random technology-specific productivity shocks and bribe requests, and endogenously choose between modern and traditional production technologies given the productivity shocks.  We use firm-level data from 219 World Bank Enterprise Surveys (WBES) to calibrate country-specific model parameters. The data contain direct reports of annual bribe payments and production decisions of about 37,000 firms operating in 148 different countries in 2006-2023.

We quantify the impact of heterogeneous bribery on capital accumulation, technology adoption, and aggregate output by conducting a series of counterfactual exercises, where we vary the bribery rates of modern and traditional firms in low-income, middle-income, and high-income countries given the country-specific calibrated parameters. The results of the counterfactual analysis support our argument that bribery can encourage adoption of more productive modern technology, foster capital accumulation, and increase aggregate output. The partial elimination of bribery (no bribery for modern firms with bribery of traditional firms kept intact) increases capital accumulation by 12.7\% in low-income countries, by 5.6\% in middle-income countries, and by 2.6\% in high-income countries. The same intervention increases the adoption of modern technology by 4.2\% in low-income countries, by 2.6\% in middle-income countries, and by 1.1\% in high-income countries. The aggregate output increases by 6.0\% in low-income countries, by 2.8\% in middle-income countries, and by 1.5\% in high-income countries (see results in Table \ref{tab:no_modern_bribe}). 

Our main result is that this partial elimination of bribery increases capital accumulation, adoption of modern technology, and aggregate output more than the complete elimination of bribery in all groups of countries, suggesting that some degree of bribery can improve aggregate economic performance. The increase in capital stock under partial elimination of bribery exceeds that under complete elimination of bribery by 18.7\% (12.7\% versus 10.7\%) in low-income countries, by 24.4\% (5.6\% versus 4.5\%) in middle-income countries, and by 44.4\% (2.6\% versus 1.8\%) in high-income countries. Similarly, the increase in aggregate output under partial elimination of bribery exceeds that under complete elimination of bribery by 3.4\% (6.0\% versus 5.8\%) in low-income countries, by 7.7\% (2.8\% versus 2.6\%) in middle-income countries and by 15.4\% (1.5\% versus 1.3\%) in high-income countries. Because the increases in capital stock and aggregate output from partial elimination of bribery are substantially larger in absolute terms in low-income countries than in middle- and high-income countries, our main result implies that anti-bribery measures selectively targeting sectors with more productive modern technologies can narrow cross-country gaps in capital stock and aggregate output.

To pinpoint the sources of increases in capital stock and aggregate output, we decompose these increases into intensive and extensive margins. The extensive margin measures the effect of the elimination of bribery due to firms' decision to adopt modern technology, and the intensive margin measures the effect of the elimination of bribery due to higher entry of firms and the more intensive utilization of inputs by incumbent firms. We find that the disproportionately large increases in capital stock and aggregate output under partial elimination of bribery relative to complete elimination of bribery are driven by larger extensive margins in both relative and absolute terms. The extensive margins under partial elimination of bribery account for about 32\% of the overall increases in aggregate output in low- and middle-income countries and for 20\% of the corresponding increase in high-income countries. The extensive margins under complete elimination of bribery account for 17.2\% of the overall increase in aggregate output in low-income countries and for 15.4\% of the corresponding increase in middle-income countries. In high-income countries, the extensive margin is negative 0.1\%, suggesting that the complete elimination of bribery discourages the adoption of modern technology in high-income countries.  

The driving force behind our main result is the higher adoption of modern technology by entrant firms that would have adopted traditional technology. When bribery is eliminated only for modern firms, as is in the case with the partial elimination of bribery, a larger fraction of entrant firms adopt more productive modern technology than when there is no bribery at all, and this raises capital stock and aggregate output above the levels when bribery is completely eliminated. These disproportionate increases in capital stock and aggregate output do not necessarily translate to a disproportionate increase in household consumption. In fact, we find that in low-income countries the overall increase in household consumption is lower under partial elimination of bribery than under complete elimination of bribery (2.8\% under partial elimination of bribery versus 3.0\% under complete elimination of bribery)\footnote{In middle- and high-income countries, the household consumption under partial elimination of bribery exceeds that under complete elimination of bribery.}. 

Our main result indicates that the aggregate output in an economy with bribery can exceed that in an economy without bribery. This may appear to violate the First Fundamental Theorem of Welfare Economics. However, since bribery introduces an element of imperfect information into firms' decision problem, the key assumption of complete information in the First Fundamental Theorem does not hold in our setting, and the implications of the theorem do not apply. 

Following the literature on corruption (e.g. \cite{ShleiferVishny1993}, \cite{FismanSvensson2007}), we model bribery as a random tax on firms' revenue. In our setting, each firm faces a random technology-specific bribe request after having entered the market and after having made an irreversible choice of production technology. Hence, each firm faces an incomplete information decision problem when making the irreversible choice of production technology. In this way, bribery generates two types of frictions in our model: the usual tax friction and the informational friction due to the uncertain nature of bribery. As \cite{DavidHopenhaynVenkateswaran2016} show, the informational friction alone can be a significant source of the firm-level resource misallocation.
 
Our findings are generally consistent with existing results in the literature: We find that the elimination of bribery increases aggregate output and capital stock, encourages the adoption of modern technology and the entry of new firms in all groups of countries, and more so in low-income countries (see results in Table \ref{tab:no_bribe})\footnote{See, for example, \cite{Mauro1995}, \cite{Ugur2014}, \cite{CieslikGoczek2018}, \cite{GrundlerPotrafke2019}.}. Our main result, however, has not been previously found in the literature, as we find that some level of bribery can foster capital accumulation and aggregate production beyond the levels without any bribery. 

Because we view bribery as a form of a tax on firms' resources, our paper is related to the literature on taxation and misallocation of resources (e.g. \cite{Restuccia2008}, \cite{HsiehKlenow2009}, \cite{Guo2019}). As in this literature, the key element in our paper that generates cross-country differences in output is the interaction between the heterogeneity of firms and policy distortions. In our model, firms differ in productivity levels, production technologies, and bribe requests. Our modeling approach, however, is different from the ones in the literature in how we model bribe requests. Following \cite{ShleiferVishny1993}, \cite{FismanSvensson2007}, \cite{Olimov2024}, we assume that bribery is distinct from taxation because bribery is secret, uncertain, and firm-specific. The uncertain aspect
of bribery connects our paper to the literature on informational frictions and resource misallocation (e.g. \cite{DavidHopenhaynVenkateswaran2016}, \cite{DavidVenkateswaran2019}).

Our empirical results are consistent with existing findings in cross-country studies of bribery and economic growth. Although existing cross-country studies generally agree on the negative relationship between bribery and economic performance (e.g. \cite{Mauro1995}, \cite{Ugur2014}, \cite{CieslikGoczek2018}, \cite{GrundlerPotrafke2019}), they find that the negative relationship breaks down or even reverses for some groups of countries. The main goal of many such studies has been to identify the characteristics of countries for which the negative relationship between bribery and growth breaks down. For example, the literature finds that the negative bribery-growth relationship breaks down for large countries in East Asia (\cite{RockBonnet2004}), for countries with weak institutions (\cite{MeonWeil2010}), or for countries with autocratic regimes (\cite{Uberti2022}). 

Our results provide an additional explanation for why the negative bribery-growth relationship may break down or even reverse. We argue that the composition of bribery rates in different sectors of an economy, rather than an average bribery rate, is crucial. For example, a country with a high average bribery rate but a low bribery rate in the modern (capital-intensive) sector can have higher capital accumulation and higher growth than a country with a low average bribery rate but a high bribery rate in the modern sector. Therefore, the reversal of the negative bribery-growth relationship in some countries can be explained by the unaccounted differences in the compositions of bribery rates that are correlated with country characteristics such as geographical location, form of government, or the rule of law. We discuss this point in the context of the East Asian paradox (\cite{Wedeman2003}, \cite{Blackburn2009}) in Section \ref{East_Asian_paradox}.

Our paper is related to the literature that relies on a dynamic general equilibrium model to study the effects of bribery. These studies include \cite{Ehrlich1999}, which focuses on the effect of bribery on human capital allocation between political capital that allows bureaucrats to extract rents (bribes) and productive capital that generates economic growth. In particular, \cite{Ehrlich1999} show that the investment in political capital is always positive, suggesting that some amount of bribery always exists in equilibrium. \cite{Sarte2000} studies the effect of bribery on the entry of firms into formal and informal economies and shows that bribery, viewed as an entry barrier to the formal economy, encourages the entry of firms into the less efficient informal economy, thus lowering economic growth. \cite{Blackburn2007} and \cite{Ivanyna2016} identify the negative effect of bribery on growth through the interaction between bribery and tax evasion. They show that when households use bribery to avoid paying income taxes, aggregate savings and capital accumulation rates are reduced (\cite{Blackburn2007}), and fewer funds are available for public investment projects (\cite{Ivanyna2016}). \cite{dagostino2016} study the negative effect of bribery on growth through complementarities between bribery and public military spending. 

Unlike \cite{Blackburn2007}, \cite{dagostino2016}, and \cite{Ivanyna2016}, we do not study the public finance aspect of bribery, and unlike \cite{Ehrlich1999} we do not study the effect of bribery on the misallocation of human capital in an economy. Our approach is similar to that of \cite{Sarte2000}, but instead of looking at how bribery affects firms' entry decisions into formal and informal economies, we focus on the effect of bribery on firms' decisions to adopt a more productive modern technology or a less productive traditional technology. In this regard, our paper is related to the literature on technology adoption in developing countries (see \cite{Verhoogen2023} for the most recent overview). 

The rest of the paper is as follows. In Section \ref{model}, we present the dynamic stochastic general equilibrium model, where firms face random bribe requests and differ in productivity draws and the choice of production technology. In Section \ref{calibration}, we discuss the data and estimated parameter values that we use in counterfactual analysis. In Section \ref{counterfactual_analysis}, we present three counterfactual scenarios to illustrate how bribery can harm and foster capital accumulation and aggregate output. In Section \ref{conclusion}, we conclude. 

\section{The model}
\label{model}

To examine how bribery influences technology adoption, we construct a dynamic stochastic general equilibrium model featuring heterogeneous firms that endogenously choose between modern and traditional production technologies. The economy consists of a representative household, a government, and a continuum of firms that differ in their technology choices, technology-specific productivity draws, and exposure to bribery. 
Upon entry, firms face stochastic productivity shocks and must make an irreversible choice of technology before the technology-specific bribe request is realized. 
The model captures heterogeneity in bribery rates across different technologies and examines the implications of bribery for aggregate technology adoption and economic performance. We proceed by outlining the decision problem faced by each economic agent and the equilibrium conditions.


\subsection{Household's problem}

The representative household lives forever and is endowed with one unit of labor in each period $t$. The household maximizes its present discounted lifetime utility given by
\begin{equation}
    \sum_{t=0}^\infty \beta^t u(C_t)
\end{equation}
where $\beta \in (0,1)$ is the discount factor and $C_t$ is consumption in period $t$. The household's budget constraint is
\begin{equation}
    C_t + K_{t+1} = r_t K_t + w_t + \Pi_t + (1-\delta) K_t + B_t, \quad \forall t
\end{equation}
where $K_t$ is the capital stock, $r_t$ is the capital rental rate, $w_t$ is the real wage, $\Pi_t$ is the total profit from the household's ownership of all operating firms, and $B_t$ is the lump-sum transfer from the government. The first-order condition for the household's optimization problem is
\begin{equation}
    u'(C_t) = \beta E_t \left\{ u'(C_{t+1}) \left[ r_{t+1} + (1-\delta) \right] \right\} \label{Euler}
\end{equation}
Equation (\ref{Euler}) represents the standard Euler equation, which equates the marginal rate of substitution of the household to the expected return on capital. The representative household’s first-order condition (\ref{Euler}) determines the steady-state equilibrium return on capital $r$ as 
\begin{equation}
    r = \frac{1}{\beta} - (1-\delta)
\end{equation}

\subsection{Firm's problem}

There are two types of firms in this model: entrants and incumbents. Entrants face a fixed entry cost to enter the market. Upon entry, each entrant observes two independent productivity shocks, one for traditional technology and one for modern technology, and makes an irreversible choice of production technology. Modern technology entails a higher operating cost than traditional technology. After choosing the production technology, each entrant learns the size of the bribe payment that remains fixed thereafter. 

Incumbents operate with production technology that they previously chose when entering the market. At the end of each period, both entrants and incumbents exit the market with an exogenous probability. The remaining entrants join the surviving incumbents in the next period.

In the following sections, we give details of the firms' decision problem and discuss how the distributions of firms' types evolve dynamically.

\subsubsection{Incumbents}
Each incumbent firm $i$ operates exclusively with traditional $(j = 0)$ or modern $(j = 1)$  technology. Both technologies exhibit decreasing returns to scale in capital and labor. The output of incumbent $i$ operating with technology $j$ is
\begin{equation}
    y_{i,j} = (A_j s_i^j)^{\sigma_j} (k_{i,j}^{\alpha_j} n_{i,j}^{1-\alpha_j})^{1-\sigma_j} \label{prod}
\end{equation}
Here, $A_j$ denotes total factor productivity, $\alpha_j$ indicates the capital input share, and $\sigma_j$ represents the span-of-control parameter reflecting managerial capacity and scalability. These parameters are common to all firms operating with technology $j$.\footnote{Differences in these parameters—total factor productivity $A_j$, capital input shares $\alpha_j$, and span-of-control parameters $\sigma_j$—drive the output differences between modern and traditional firms. These parameters are directly estimated from the data.} Modern technology is characterized by greater capital intensity, implying $\alpha_1(1-\sigma_1) > \alpha_0(1-\sigma_0)$.

Firms with the same technology \( j \) differ in two key dimensions: their idiosyncratic productivity draw \( s_i^j \) and the size of the technology-specific bribe request \( \tau_{i,j} \). The productivity draw \( s_i^j \) captures the firm-level efficiency and is assumed to follow a distribution specified later. Following \cite{Restuccia2008}, we assume that \( s_i^j \) remains fixed over time once realized. The bribe request \( \tau_{i,j} \), which is modeled as a firm-specific sales tax (see \cite{ShleiferVishny1993}; \cite{FismanSvensson2007}), denotes the fraction of output the firm must pay as a bribe to continue operating. Similarly to the productivity draw, the bribe request is drawn from a technology-specific distribution that will be described in detail later.

Each firm with technology $j$ incurs a fixed technology-specific operating cost \( c_j \) that must be paid every period. Operation with modern technology entails a higher operating cost, with \( c_0 = 0 \) for a firm with traditional technology and \( c_1 > 0 \) for a firm with modern technology. Given these parameters, the incumbent \( i \) with technology \( j \) chooses capital \( k_{i,j} \) and labor \( n_{i,j} \) to maximize its current-period profit:
\begin{equation}
    \pi_{j}(s_i^j, \tau_{i,j})= (1 - \tau_{i,j}) y_{i,j} - r k_{i,j} - w n_{i,j} - c_j,
\end{equation}
where \( r \) is the rental rate of capital and \( w \) is the real wage.




Assuming perfectly competitive factor markets, the firm’s optimal capital and labor input choices, $k_{j}(s_i^j, \tau_{i,j})$ and $n_{j}(s_i^j,\tau_{i,j})$, depend on technology-specific productivity draws and bribe requests and are as follows: 
\begin{equation}
    k_{j}(s_i^j, \tau_{i,j}) = ((1-\sigma_j)(1-\tau_{i,j}))^{\frac{1}{\sigma_j}} A_j s_i^j \left(\frac{\alpha_j}{r}\right)^{\frac{1-(1-\alpha_j)(1-\sigma_j)}{\sigma_j}} \left(\frac{1-\alpha_j}{w}\right)^{\frac{(1-\alpha_j)(1-\sigma_j)}{\sigma_j}}
\end{equation}
and 
\begin{equation}
    n_{j}(s_i^j,\tau_{i,j}) = ((1-\sigma_j)(1-\tau_{i,j}))^{\frac{1}{\sigma_j}} A_j s_i^j \left(\frac{\alpha_j}{r}\right)^{\frac{\alpha_j(1-\sigma_j)}{\sigma_j}} \left(\frac{1-\alpha_j}{w}\right)^{\frac{1-\alpha_j(1-\sigma_j)}{\sigma_j}}.
\end{equation}
Then, the optimal level of output $y_{j}(s_i^j, \tau_{i,j})$ of the incumbent $i$ is
\begin{align}
    y_{j}(s_i^j, \tau_{i,j}) = (1-\tau_{i,j})^{\frac{1-\sigma_j}{\sigma_j}} A_j s_i^j h_{j},
\end{align}
where $h_{j}$ captures the impact of factor prices and technology parameters common to all firms with technology $j$ and is defined as follows:
\begin{equation}
    h_{j} \equiv (1-\sigma_j)^{\frac{1-\sigma_j}{\sigma_j}} \left(\frac{\alpha_j}{r}\right)^{\frac{\alpha_j(1-\sigma_j)}{\sigma_j}} \left(\frac{1-\alpha_j}{w}\right)^{\frac{(1-\alpha_j)(1-\sigma_j)}{\sigma_j}}
\end{equation}

Since all incumbents with traditional technology have zero operating cost (\( c_0 = 0 \)), the maximum current-period profit of the incumbent \( i \) operating with traditional technology (\( j = 0 \)) and facing the realized bribe request \( \tau_{i,0} \) is given by
\[
\pi_{0}(s_i^0, \tau_{i,0}) = \sigma_0 (1 - \tau_{i,0})^{\frac{1}{\sigma_0}} A_0 s_i^0 h_0
\]

In contrast, since all incumbents with modern technology incur a positive fixed operating cost (\( c_1 > 0 \)) each period, the maximum current-period profit of the incumbent \( i \) operating with modern technology (\( j = 1 \)) and facing the realized bribe request \( \tau_{i,1} \) is
\[
\pi_{1}(s_i^1, \tau_{i,1}) = \sigma_1 (1 - \tau_{i,1})^{\frac{1}{\sigma_1}} A_1 s_i^1 h_1 - c_1
\]


Finally, given that incumbents exit the market with an exogenous probability \( \lambda \) every period, the incumbent $i$'s present discounted value of operating with traditional technology $(j=0)$ in period $t$ is defined recursively as
\[
W_{0,t}(s_i^0, \tau_{i,0}) = \pi_{0}(s_i^0, \tau_{i,0}) + E_t \left[\frac{1-\lambda}{1+R} W_{0,t+1}(s_i^0, \tau_{i,0})\right],
\]
and the incumbent $i$'s present discounted value of operating with modern technology $(j=1)$ in period $t$ is defined recursively as
\[
W_{1,t}(s_i^1, \tau_{i,1}) = \pi_{1}(s_i^1, \tau_{i,1}) + E_t \left[\frac{1-\lambda}{1+R} W_{1,t+1}(s_i^1, \tau_{i,1})\right],
\]
where $R = r - \delta$ represents the net real interest rate.

\subsubsection{Entrants}

There is an infinite continuum of potential entrants. Each potential entrant decides whether to enter the market and faces a one-time fixed cost of entry \( c_e \). Prior to entry, potential entrants observe only the distributions of technology-specific productivity draws and technology-specific bribe requests. Upon entry, each entrant obtains two productivity draws \( \{s_i^0, s_i^1\} \) independently from two identical Fréchet distributions:
\[
F(s_i^j) = \exp\left[-\left(\frac{s_i^j}{\phi}\right)^{-\theta}\right], \quad \text{where } \phi \text{ is chosen such that } E[s_i^j] = 1.
\]
The shape parameter \( \theta \), referred to as the ``technology elasticity" in \cite{Farrokhi2024}, governs the dispersion of productivity draws. The realized productivity draws remain fixed over time.

After observing productivity draws, but before learning the actual bribe requests, entrants choose between traditional (\( j = 0 \)) and modern (\( j = 1 \)) technologies. In this stage, entrants only know the distributions of technology-specific bribe requests \( P_j(\tau_{i,j}) \). Bribe requests follow the Bernoulli distribution with support \( \tau_{i,j} \in \{0, \bar{\tau}_j\} \). Specifically, the probability of facing a bribe request of size \( \bar{\tau}_j \) is $p_j=P_j(\tau_{i,j}=\bar{\tau}_j)$, and the probability of not facing a bribe request is $1-p_j=P_j(\tau_{i,j}=0)$. Both probability \( p_j \) and bribe request \( \bar{\tau}_j \) depend on technology \( j \), reflecting systematic differences in the corruption environment in sectors with different production technologies.\footnote{Although we assume that bribery differentially affects modern and traditional firms, we do not make any assumptions about whether bribery more heavily affects modern or traditional firms. We use data to determine how much bribery affects modern firms compared to traditional firms in different countries. The existing literature does not give a decisive answer to whether modern or traditional firms pay larger bribes, and the available results differ by country. For example, \cite{Svensson2003} finds that (modern) firms with a higher capital-labor ratio in Uganda tend to pay higher bribes than (traditional) firms with a lower capital-labor ratio, because a higher capital intensity is correlated with higher future expected profitability of firms and therefore higher bribe demands from Ugandan bureaucrats. On the other hand, \cite{BaiJayachandranMaleskyOlken2019} find that larger Vietnamese firms with operations in multiple provinces tend to pay lower bribes because such firms can more flexibly shift operations between branches in different provinces and, therefore, can negotiate lower bribe payments to provincial bureaucrats. Since larger firms also tend to be more capital-intensive and operate with modern technology, this line of argument would imply a lower bribery rate for modern firms.}

Once the entrant $i$ adopts the technology $j$, the uncertainty about the bribe request is resolved, and the government extracts the realized bribe payment \( \tau_{i,j} \in \{0, \bar{\tau}_j\} \) from the firm every period. Since the bribe represents a constant share of the firm’s revenue paid to the government, we assume that \( \bar{\tau}_j \leq 1 \). 

Although we model a bribe request as a sales tax, there are three fundamental differences between a bribe in our model and a conventional sales tax, all arising from the illicit and secret nature of bribery (e.g. \cite{ShleiferVishny1993}; \cite{Olimov2024}). First, to capture the secrecy of bribery, we assume that firms do not observe actual bribe requests when making irreversible technology choices. This contrasts with taxation, where statutory tax rates are publicly known and factored into firms’ decisions ex ante. Second, because bribe payments are privately negotiated, even firms with identical technologies and sales levels may face different bribe requests. To capture this arbitrariness of bribe requests, we model the bribe request as a random variable. This is different from standard tax policy, where firms with identical characteristics are subject to similar tax rates. Third, to capture systematic differences in bribery rates in sectors with traditional (labor-intensive) and modern (capital-intensive) production technologies, the distributions of the bribe requests differ depending on the firm's choice of technology \(j\). This technology-specific risk from bribery introduces an additional layer of distortion that is absent in standard taxation.

Given the timing described above, the entrant \( i \) with realized productivity draws \( \{s_i^0, s_i^1\} \) chooses technology \( j \in \{0,1\} \) to maximize the expected lifetime profit, taking the expectation over the distribution of technology-specific bribe requests. Let \( \hat{W}_{j,t}(s_i^j) \) denote the expected lifetime payoff of the entrant $i$ from the adoption of technology \( j \) in period \( t \). Then, the expected lifetime payoff from the adoption of traditional technology in period $t$ is

\[\hat{W}_{0,t}(s_i^0) = (1 - p_0) W_{0,t}(s_i^0, 0) + p_0 W_{0,t}(s_i^0, \bar{\tau}_0)\]

and from the adoption of modern technology in period $t$ is 

\[
\hat{W}_{1,t}(s_i^1) = (1 - p_1) W_{1,t}(s_i^1, 0) + p_1 W_{1,t}(s_i^1, \bar{\tau}_1)
\] 

The entrant $i$ adopts the technology $j$ that gives a higher expected lifetime payoff:
\[
j = 
\begin{cases}
1 & \text{if } \hat{W}_{1,t}(s_i^1) \geq \hat{W}_{0,t}(s_i^0), \\
0 & \text{otherwise}.
\end{cases}
\]
This decision rule yields the threshold productivity in the modern sector $\bar{s}^1(s_i^0)$ as a function of $s_i^0$, where
\[
\bar{s}^1(s_i^0) \equiv \hat{W}_{1,t}^{-1} \left(\hat{W}_{0,t}(s_i^0)\right)
\]
The entrant $i$ adopts modern technology ($j=1$) whenever its realized productivity draw with modern technology $s_i^1$ exceeds the threshold level $\bar{s}^1(s_i^0)$, or $s_i^1 \geq \bar{s}^1(s_i^0)$.



Let $\eta_t$ denote the share of entrants adopting modern technology in period $t$. Then, given the threshold level $\bar{s}^1(s_i^0)$, the distributions of technology-specific productivity draws $F(s^j)$, and the distributions of bribe requests, the share of entrants adopting modern technology in period $t$ is  
\begin{align}
\eta_t &= \int_0^{\infty} \int_{s_i^1 \geq \bar{s}^1(s_i^0)  }  ^{\infty} dF(s^0)dF(s^1) 
\end{align}

The share of entrants adopting modern technology $\eta_t$ is determined by: $(i)$ technology-specific productivity parameters ($\sigma_j$, $A_j$, $\alpha_j$, $F(s^j)$), $(ii)$ technology-specific operating cost ($c_1$), $(iii)$ technology-specific distributions of bribe requests ($P_0(\tau),P_1(\tau)$), $(iv)$ equilibrium factor prices ($w$, $r$), and $(v)$ the shape parameter $\theta$ that influences firms' technology adoption decisions.\footnote{In the stationary equilibrium, the share of entrants adopting the modern technology is given by \begin{align}
\eta &= \int_0^{\infty} \int_{s^1 \geq C's^0+D'}^{\infty} dF(s^0)dF(s^1) \notag \\
  &= \int_0^{\infty} [1-F(C's^0+D')] dF(s^0)\notag  \\
  &= 1 - \int_0^{\infty} e^{-(\frac{C's^0+D'}{\phi})^{-\theta}}\frac{\theta}{\phi}(\frac{s^0}{\phi})^{-(1 + \theta)}e^{-(\frac{s^0}{\phi})^{-\theta}}ds^0
\end{align}
where \(C' \equiv \frac{\sigma_0A_{0}h_{0}E_{T_0}[(1-T_0)^\frac{1}{\sigma_{0}}]}{\sigma_1A_1h_1E_{T_1}[(1-T_1)^\frac{1}{\sigma_1}]}\) and \(D' \equiv \frac{c_1}{\sigma_1A_1h_1E_{T_1}[(1-T_1)^\frac{1}{\sigma_1}]}\).
}

Since the entrant $i$ knows the distributions of technology-specific productivity draws, $F(s^j)$, and the expected benefits from the adoption of traditional and modern technologies, $\hat{W}_{0,t}(s_i^0)$  and $\hat{W}_{1,t}(s_i^1)$, the entrant $i$ can calculate the expected payoff of entry net of the entry cost $c_e$. If the expected net payoff of entry is positive, the entrant $i$ enters the market in the period $t$. We denote the expected net payoff of entry by $W_{i,t}^e$, where
\[
    W_{i,t}^e = -c_e + \int_0^{\infty} \int_0^{\infty} \max\{ \hat{W}_{0,t}(s_i^0), \hat{W}_{1,t}(s_i^1)\} dF(s_i^0) dF(s_i^1) 
\]

Since all entrants are ex ante identical, the expected net payoff of entry $W_{i,t}^e$ is identical for all entrants, and $W_{i,t}^e=W_t^e$, $\forall i$. Hence, an entrant that is indifferent between entering and not entering has $W_t^e$ equal zero. This gives the following free entry condition for entrants in each period $t$: 
\begin{equation}
    W_t^e=0
\label{eq:free_entry}
\end{equation}

\subsection{Laws of motion}
Let \(\mu_{j,t}(s^j,\tau_{j})\) denote the joint distribution of productivity draws and bribe requests of all firms operating with technology \(j\) in period \(t\). With $m_t$ representing the total number of entrants and $\eta_t$ representing the share of entrants adopting modern technology in period $t$, we obtain the following laws of motion for the distributions of firms operating with different technologies and facing different bribe requests in period \(t+1\):
\begin{align}
    \mu_{0,t+1}(s^0,\tau_{0}) &= (1-\lambda)\mu_{0,t}(s^0,\tau_{0}) + m_t (1-\eta_t) P_0(\tau_0)F(s^0)
    \label{eq:motion_0}
\end{align}
and
\begin{align}
    \mu_{1,t+1}(s^1,\tau_{1}) &= (1-\lambda)\mu_{1,t}(s^1,\tau_{1}) + m_t \eta_t P_1(\tau_1)F(s^1)
    \label{eq:motion_1}
\end{align}
The first terms in both equations represent the mass of incumbents operating with each technology type that do not exit the market in period $t$. The second terms in both equations represent the mass of entrants who choose to adopt a particular technology in period $t$. The mass of entrants adopting traditional technology in period $t$ is \(m_t (1 - \eta_t)\), and the mass of entrants adopting modern technology in period $t$ is \(m_t \eta_t\).

\subsection{The government}


The government balances its budget in every period $t$. Specifically, the government collects bribes from firms operating with both technologies and redistributes all bribery proceeds to the household as a lump sum transfer, while maintaining a balanced budget. The government's budget constraint is given by
\begin{equation} B_t = \int \tau_0 y_{0,t}(s^0,\tau_0) d\mu_{0,t}(s^0,\tau_0) + \int \tau_1 y_{1,t}(s^1,\tau_1) d\mu_{1,t}(s^1,\tau_1), \label{eq
} \end{equation} where $y_{j,t}(s^j,\tau_j)$ is the output of a firm operating with technology $j$ in period $t$ and \(B_t\) represents the total lump sum transfer to the household. The lump sum transfer $B_t$ is equal to the sum of all bribery proceeds from firms operating with traditional and modern technologies.

\subsection{Market-clearing conditions}
The market-clearing conditions for labor and capital require that the total factor demand of firms operating with traditional and modern technologies is equal to the total factor supply of the household. For the labor input, this condition is 
\begin{equation}
    1 = N_t = \int n_{0,t}(s^0,\tau_0) d\mu_{0,t}(s^0,\tau_0) + \int n_{1,t}(s^1,\tau_1) d\mu_{1,t}(s^1,\tau_1),
    \label{eq:N_t}
\end{equation}
where \(N_t\) is the total supply of labor and the right-hand side represents the total demand for labor of firms operating with traditional and modern technologies, respectively.

The market-clearing condition for the capital input is
\begin{equation}
    K_t = \int k_{0,t}(s^0,\tau_0) d\mu_{0,t}(s^0,\tau_0) + \int k_{1,t}(s^1,\tau_1) d\mu_{1,t}(s^1,\tau_1),
    \label{eq:K_t}
\end{equation}
where \(K_t\) denotes the total supply of capital, and the right-hand side represents the demand for capital of firms operating with traditional and modern technologies, respectively.

The aggregate output of the economy combines the output of all firms and is given by
\begin{equation}
    Y_t = \int y_{0,t}(s^0,\tau_0) d\mu_{0,t}(s^0,\tau_0) + \int y_{1,t}(s^1,\tau_1) d\mu_{1,t}(s^1,\tau_1),
    \label{eq:Y_t}
\end{equation}
where the right-hand side represents the total output of firms operating with traditional and modern technologies, respectively.

The total number of operating firms in the economy in period \(t\) is given by
\begin{equation}
    M_t = \int d\mu_{0,t}(s^0,\tau_0) + \int d\mu_{1,t}(s^1,\tau_1),
\end{equation}
which represents the total mass of firms operating with traditional and modern technologies. The fraction of all (entrants and incumbents) firms operating with modern technology in period $t$ is 
\begin{equation}
    \rho_t = \frac{\int d\mu_{1,t}(s_1,\tau_1)}{\int d\mu_{0,t}(s_0,\tau_0) + \int d\mu_{1,t}(s_1,\tau_1)}
\end{equation}

Finally, the aggregate resource constraint captures how the total output is allocated between consumption, capital investment, operating costs, and it has the following form:
\begin{equation}
    C_t + K_{t+1} + c_e m_t + c_1 \rho_t M_t = Y_t + (1 - \delta) K_t
\end{equation}
This constraint indicates that the current-period aggregate output \(Y_t\) covers the current-period household's consumption \(C_t\), investment in the next-period capital \(K_{t+1}-(1 - \delta) K_t\), entry costs of the current-period entrants \(c_e m_t\), and operating costs of all modern firms \(c_1 \rho_t M_t\).

\subsection{Stationary equilibrium}

Given our view of bribery as a relatively stable, time-invariant distortion, we abstract from the aggregate risk and concentrate on a stationary equilibrium in which all aggregate variables remain constant over time. Following \cite{Restuccia2008}, we characterize a stationary competitive equilibrium that features time-invariant distributions of operating firms with different productivity draws and technology-specific bribe requests. The steady-state equilibrium values are denoted without time subscripts.

The steady-state equilibrium consists of factor demands \(\{n_j(s^j, \tau_j), k_j(s^j, \tau_j)\}\), input prices \(\{w, r\}\), time-invariant joint distributions of operating firms with different productivity draws and bribe requests \(\mu_j(s^j, \tau_j)\), the share of firms operating with modern technology \(\rho\), the number of entrants \(m\), the total number of operating firms in the economy $M$, aggregate household consumption \(C\), aggregate capital \(K\), and the lump sum aggregate bribe transfer \(B\), such that:
\begin{enumerate}[i.]
    \item The representative household maximizes utility subject to the budget constraint;
    \item Firms choose \(\{n_j(s^j, \tau_j), k_j(s^j, \tau_j)\}\) to maximize profits given input prices \(\{w, r\}\);
    \item The free entry condition holds: \(W^e = 0\);
    \item The government balances its budget:
    \begin{equation}
        B = \int \tau_0 y_0(s^0, \tau_0) d\mu_0(s^0, \tau_0) + \int \tau_1 y_1(s^1, \tau_1) d\mu_1(s^1, \tau_1);
    \end{equation}
    \item The markets for labor and capital inputs clear:
    \begin{equation}
        1 = N = \int n_0(s^0, \tau_0) d\mu_0(s^0, \tau_0) + \int n_1(s^1, \tau_1) d\mu_1(s^1, \tau_1),
    \end{equation}
    \begin{equation}
        K = \int k_0(s^0, \tau_0) d\mu_0(s^0, \tau_0) + \int k_1(s^1, \tau_1) d\mu_1(s^1, \tau_1);
    \end{equation}
    \item The aggregate resource constraint holds: \begin{equation}
        Y = \int y_0(s^0, \tau_0) d\mu_0(s^0, \tau_0) + \int y_1(s^1, \tau_1) d\mu_1(s^1, \tau_1) = C + \delta K + c_e m + c_1 \rho M; \end{equation}
    \item The joint distributions of operating firms with different productivity draws and bribe requests are time-invariant:
    \begin{equation}
        \mu_j(s^j, \tau_j) = \frac{m}{\lambda}(1-\eta) P_j(\tau_j)F(s^j), \quad j \in \{0,1\}
    \end{equation}
\end{enumerate}

\section{Quantitative analysis}
\label{calibration}
To quantify the impact of heterogeneous bribery on technology adoption and aggregate output, we use firm-level data from World Bank Enterprise Surveys (WBES) to calibrate country-specific model parameters. 
Section \ref{Parameter_estimation} describes the data and our calibration procedure. Section \ref{Data_patterns} details the key patterns in the data along with the calibrated parameter values. 
Lastly, in Section \ref{East_Asian_paradox}, we discuss our results in the context of the ``East Asian paradox.''


\subsection{Data and parameter calibration}
\label{Parameter_estimation}

We use firm-level data from the World Bank Enterprise Surveys (WBES) to calibrate parameters of our theoretical model. These surveys contain responses from business owners and senior managers about their firms' operational decisions and experiences with bureaucratic environments in multiple countries. 
Given that we have data from multiple surveys for each country, we treat data from each survey as the data from a unique observational economy.\footnote{For instance, we consider firms surveyed in China in 2007 separately from those surveyed in China in 2023. We do not combine survey data from the same country because we do not have unique firm-level identifiers.} Since survey responses were collected in local currencies and in different years, we convert all values into US dollars at 2009 constant prices. In addition, we eliminate observations with outliers and with missing values\footnote{We use the three-standard-deviation rule to identify outliers. Outliers are identified based on values in variables d2 (sales), n7a (capital), n2a (cost of labor), and n2e (costs of intermediate input) in the questionnaire.}. In total, our data include 219 firm-level WBES surveys of approximately 37,000 firms operating in 2006-2023.

Following the methodologies of \cite{Lloyd1982} and \cite{Farrokhi2024}, we use the k-means clustering algorithm to classify firms in each economy into ``modern" and ``traditional" groups. A firm $\omega$ is classified as modern if its capital-labor ratio $\kappa(\omega)$ exceeds a country-specific threshold $\kappa(\omega^*)$ and is classified as traditional otherwise. The threshold minimizes within-group variance of capital-labor ratios and is formally defined as
\[
\omega^* = \arg\min \left[\sum_{\omega \in \Omega_0} (\kappa_0(\omega)-\bar{\kappa}_0)^2 + \sum_{\omega \in \Omega_1} (\kappa_1(\omega)-\bar{\kappa}_1)^2\right],
\]
where $\bar{\kappa}_j$ is the average capital-labor ratio for firms operating with technology $j$ in each country. Here, \(\Omega_0 = \{\omega|\kappa(\omega) <\kappa(\omega^*)\}\) is the set of traditional firms in each country and \(\Omega_1 = \{\omega|\kappa(\omega) >\kappa(\omega^*)\}\) is the set of modern firms in each country. 

Given our classification results, we estimate the labor elasticity of output for each technology $j$ by using the control function approach of \cite{Levinsohn2003} and \cite{Olley1996})\footnote{To resolve the endogeneity issue, we estimate coefficients of transitory inputs by controlling for non-transitory inputs and intermediate inputs.}. Specifically, for each technology $j$ and for all countries in our data, we assume that the total sales \(y'_j\) are a linear function of the total payments to labor \(n'_j\) and a polynomial expansion of the total payments to capital \(k'_j\) and intermediate inputs \(m'_j\):
\begin{equation}
   y'_j = \hat\gamma_j n'_j + f_j(k'_j, m'_j),
\end{equation}
where $f_j(k'_j, m'_j) = k'_j \times m'_j + k'_j + m'_j + {k'_j}^2 + {m'_j}^2 + \dots + {k'_j}^5 + {m'_j}^5$ includes the interaction term and higher-order terms up to the fifth order\footnote{While including higher order expansions increases accuracy, we follow \cite{Farrokhi2024} and include terms only up to the fifth order.}. The estimated labor elasticity of output is \( \hat{\gamma}_0 = 0.479 \) for traditional firms and \( \hat{\gamma}_1 = 0.308 \) for modern firms. In our model, \( \gamma_1 \) corresponds to \( (1 - \alpha_1)(1 - \sigma_1) \) and \( \gamma_0 \) corresponds to \( (1 - \alpha_0)(1 - \sigma_0) \). We estimate the technology-specific Lucas span-of-control parameters $\sigma_j$ directly from the average profits of firms operating with technology $j$. We obtain estimates of the relative shares of capital to labor inputs $\alpha_j$ by interacting estimates of $\sigma_j$ and $1-\gamma_j$. The estimates for $\alpha_j$ and $\sigma_j$ in Table \ref{tab:calibration_summary} indicate that modern technology is indeed more capital intensive than traditional technology, $\alpha_1 (1-\sigma_1) > \alpha_0 (1-\sigma_0)$. Note that these estimates are common to all countries in our data. 

We then calibrate country-specific parameters, including entry costs \( c_e \), operating costs with modern technology \( c_1 \), total factor productivity ratios of modern and traditional firms \( \frac{A_1}{A_0} \), and distributions of bribe requests \( P_j(\tau_j) \). We jointly calibrate these parameters by matching each country's GDP per capita, the probability of encountering a bribe request for each technology, the average share of sales paid as bribes for each technology, the share of modern firms, and the output share of modern firms in the data to their corresponding theoretical predictions.

We use the normalized distribution of GDP per capita to calibrate total factor productivity ratios \( \frac{A_1}{A_0} \). We set \( A_1 = 1 \) for all countries in our data and set the lowest calibrated total factor productivity ratio \( \frac{A_1}{A_0} \) equal to unity. This normalization allows cross-country comparisons while controlling for differences in the absolute values of the per capita income across countries.\footnote{The distribution of the normalized GDP per capita differs from the that of the actual GDP per capita only in scale; otherwise, the actual and normalized distributions are identical.}

To estimate the distributions of country-specific and technology-specific bribe requests $P_j(\tau_j)$, we assume that bribe requests follow a Bernoulli process: Each firm with technology $j$ faces either a positive bribe request $\tau_j \in (0, 1]$ or no bribe request at all. The probability that a firm faces a positive bribe request is indicated by $P_j(\tau_j)$. We use the proportion of firms operating with technology $j$ that report positive bribe payments in each country to estimate these probabilities. Specifically, we use firms' responses regarding informal payments made to expedite bureaucratic processes such as customs, taxation, licensing, and regulatory compliance.\footnote{The exact wording of the corresponding question in the WBES survey is: ``It is said that establishments are sometimes required to make gifts or informal payments to public officials to `get things done' regarding customs, taxes, licenses, regulations, services, etc. On average, what percentage of total annual sales---or estimated total annual value---do establishments like this one pay in informal payments or gifts to public officials for this purpose?''
}
Specifically, country-specific and technology-specific bribe requests $\tau_j$ are calculated as weighted averages of positive bribe payments reported by firms operating with technology $j$ in each country.\footnote{Detailed country-specific and technology-specific probabilities are reported in Table \ref{tab:appendix_data_summary} in the Appendix.}


Finally, several parameters are kept constant across all countries: the discount factor $\beta = 0.96$  reflecting the annual real interest rate of 4\%, the capital depreciation rate $\delta = 0.08$, and the exit rate of firms $\lambda = 0.10$. Consistently with \cite{Farrokhi2023}, we set the shape parameter of the Fréchet distribution at $\theta = 4.5$. As our analysis focuses on steady-state equilibria, we do not explicitly specify the household utility function.\footnote{A specific utility function would be required for welfare analysis.} Table~\ref{tab:calibration_summary} summarizes all parameters.

\begin{table}[h]
\small
    \caption{Model parameters.}
    \centering
    \begin{tabular}{lcc}
        \hline
        \hline
        {Description} & {Parameter} & {Value/Source} \\
        \hline
        Time preference & \(\beta\) & 0.96 (\cite{Restuccia2008}) \\
        Depreciation rate & \(\delta\) & 0.08 (\cite{Restuccia2008}) \\
        Firms' exit rate & \(\lambda\) & 0.10 (\cite{Restuccia2008}) \\
        Technology elasticity & \(\theta\) & 4.5 (\cite{Farrokhi2023}) \\
        Productivity params. (traditional) & $\{\sigma_0,\alpha_0\}$ & \{0.378,0.230\} (WBES) \\
        Productivity params. (modern) & $\{\sigma_1,\alpha_1\}$ & \{0.334,0.538\} (WBES) \\
        TFP ratios & $\frac{A_1}{A_0}$ & Country-specific estimates (WBES) \\
        Entry costs & $c_e$ &  Country-specific estimates (WBES)\\
        Operating costs of modern firms & $c_1$ & Country-specific estimates (WBES)\\
        Distributions of bribe requests & $\{P_0(\tau_0),P_1(\tau_1)\}$ & Country-specific estimates (WBES)\\
        \hline
        \hline
    \end{tabular}
    \label{tab:calibration_summary}
\end{table}

\subsection{Data patterns and the estimated parameters}
\label{Data_patterns}

\begin{figure}[ht]
\begin{center}
\includegraphics[width=0.9\textwidth]{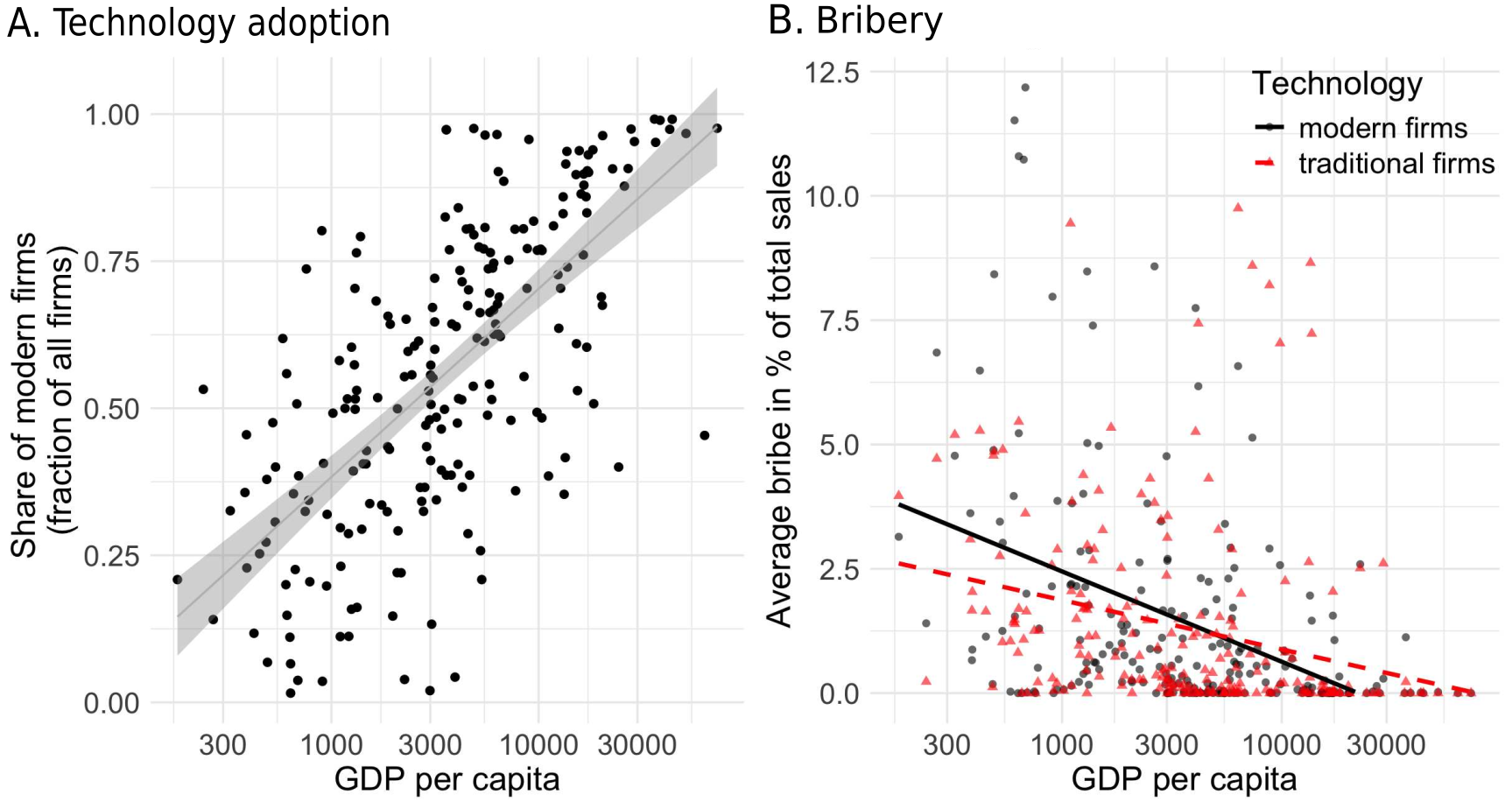}
\end{center}
\caption{Technology adoption and bribery across countries. 
Panel A: Share of modern firms (in percent of all firms) across countries. The linear fit is indicated by the solid line, and the 95\% confidence area is indicated by the gray shaded area. Panel B: Bribery rates of modern and traditional firms across countries. The vertical axis measures the share of annual sales paid as bribes in percent of annual sales by modern (circular markers) and by traditional (triangular markers) firms in each country in the data. The solid line indicates the linear fit for modern firms across all countries, and the dashed line indicates the linear fit for traditional firms across all countries. The horizontal axis in both panels measures GDP per capita in constant 2009 US dollars.}
 \label{fig:figure1}
\end{figure}

We plot the shares of modern firms for all countries in our data in panel A of Figure \ref{fig:figure1}\footnote{We report the classification results for all countries in our data in Table \ref{tab:appendix_data_summary} in the Appendix.}. In the figure, we observe that wealthier countries have higher shares of firms operating with more productive (capital-intensive) modern technology. This data pattern is consistent with results in the existing literature that also report the higher rate of adoption of capital-intensive technology in wealthier countries (e.g. \cite{Farrokhi2024}) and the increased use of capital input by more productive firms (e.g. \cite{Ciccone2002}, \cite{MidriganXu2014}, \cite{Lagakos2016}).

We report linear trend lines and scatter plots for country- and technology-specific bribe requests in panel B of Figure \ref{fig:figure1}. In the figure, each country’s bribe payments by traditional firms are indicated by triangular markers, while those by modern firms are indicated by circular markers. Our estimates indicate that in wealthier countries, firms pay smaller bribes and are less likely to face bribe requests. In addition, the different trend lines for bribe payments of traditional and modern firms across countries in panel B of Figure \ref{fig:figure1} indicate that bribery has a differential effect on firms with different production technologies in different countries. To further illustrate this point, we classify countries into three income groups and report technology-specific bribe payments, the probabilities of bribe requests, and the shares of modern firms within each income group in Table \ref{tab:summary_stats}. 

Two patterns emerge in the data. First, average bribe payments of modern and traditional firms decline with countries' income, which means that wealthier countries generally experience less bribery in both modern and traditional sectors. Second, the rates of decline in bribery differ across technologies as countries' income increases. In poorer countries, modern firms face higher bribery rates than traditional firms. The bribery rate for modern firms declines faster than for traditional firms as countries' income level increases. In high-income countries, bribery affects mainly traditional firms. 

\begin{table}[h]
        \begin{center}
        \small
        \caption{Technology adoption and bribery across countries.}
        \label{tab:summary_stats}
        \begin{tabular}{lccc}
            \hline
            \hline
            & {Low-income} & {Middle-income} & {High-income} \\
            & {countries} & {countries} & {countries} \\
            \hline 
            {Share of modern firms (\% of all firms) } & 36.8\%  & 55.4\% & 75.6\% \\
            {Bribe of traditional firms (\% of annual sales)}& 1.96\% & 0.99\% & 1.01\% \\           
            {Probability of bribe payment of traditional firms} & 28\% & 15\% & 13\% \\            
            {Bribe of modern firms (\% of annual sales)} & 2.60\% & 1.11\% & 0.60\% \\
            {Probability of bribe payment of modern firms} & 34\% & 16\% & 11\% \\
            \hline
            \hline
        \end{tabular}
        \medskip
    \footnotesize
    \end{center}
    \textbf{Notes:} The table reports the shares of modern firms and the distributions of the average bribe payments of modern and traditional firms in low-, middle- and high-income countries. Low-income countries are countries with GDP per capita between 180.6 and 2,095.8 constant 2009 US dollars (lowest third in the data). Middle-income countries are countries with GDP per capita between 2,169 and 5,848 constant 2009 US dollars (middle third in the data). High-income countries are countries with GDP per capita between 5,936 and 72,735 constant 2009 US dollars (highest third in the data).        
\end{table}


\begin{figure}[ht]
\begin{center}
\includegraphics[width=0.9\textwidth]{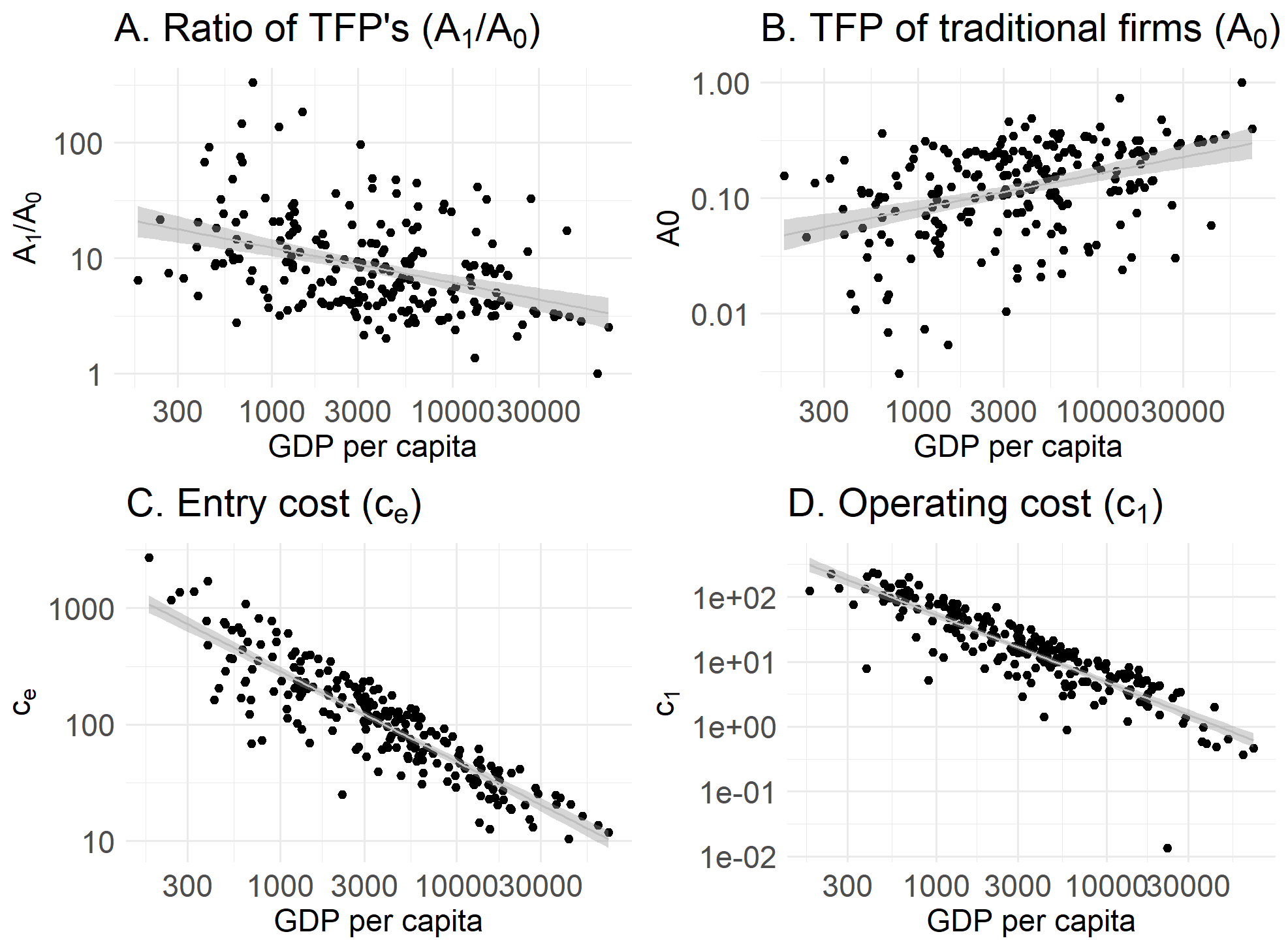}
\end{center}
 \caption{Calibrated TFP's, entry and operating costs across countries. The horizontal axis measures GDP per capita. Each dot represent a country in our data. Solid lines represent linear trends, and shaded areas indicate the 95\% confidence areas. Panel A: Calibrated ratios of total factor productivities $\frac{A_1}{A_0}$ across countries. Panel B: Calibrated total factor productivity of traditional firms across countries. Panel C: Calibrated entry costs \(c_e\) across countries. Panel D: Calibrated operating costs of modern technology \(c_1\) across countries.}
 \label{fig:calibration_results}
\end{figure}

We report calibrated country-specific estimates of $\frac{A_1}{A_0}$, $A_0$, $c_e$ and $c_1$ in Figure \ref{fig:calibration_results}. Panel A in Figure \ref{fig:calibration_results} presents the linear trend line and the scatter plot for the calibrated ratio of total factor productivities in the traditional and modern sectors, while panel B in Figure \ref{fig:calibration_results} presents the linear trend line and the scatter plot for the calibrated total factor productivity of traditional firms across countries in the data. Since $A_1=1$ for all countries in the data, these figures contain the same information and indicate that traditional firms are more productive in wealthier countries. This result is consistent with the established view in the literature that wealthier countries generally have more productive firms (e.g. \cite{Jones2016}, \cite{HallJones1999}). The patterns in panels A and B in Figure \ref{fig:calibration_results} are consistent with the assumption that firms operating with different technologies in wealthier countries have a narrower productivity gap due to higher overall efficiency.  

Panel C in Figure \ref{fig:calibration_results} reports the linear trend line and the scatter plot for calibrated entry costs ($c_e$) for firms in different countries, and panel D in Figure \ref{fig:calibration_results} reports the linear trend and the scatter plot for calibrated operating costs of modern technology ($c_1$) for firms in different countries. The patterns in both panels indicate that firms in wealthier countries have lower entry costs and modern firms in wealthier countries have lower operating costs. In particular, these patterns indicate that in wealthier countries it is not only cheaper to establish a firm, but it is also cheaper to operate a firm that utilizes advanced machinery. These patterns are consistent with existing findings in the literature that find higher entry and operating costs for firms in poorer countries (e.g. \cite{Buera2021}, \cite{BentoRestuccia2017}, \cite{ColeGreenwoodSanchez2016}, \cite{ParentePrescott1999}). 

The data patterns in Figure \ref{fig:figure1} and Table \ref{tab:summary_stats} suggest that differences in the bribery rates of modern and traditional firms may explain variations in technology adoption, capital accumulation, and aggregate output between countries. In low-income countries, where only 36.8\% of the firms are modern, the ratio of bribery rates of traditional and modern firms is 0.7. In middle-income countries, where about 55.4\% of the firms are modern, the ratio of bribery rates of traditional and modern firms is close to one. In high-income countries, where 75.6\% of the firms are modern, the ratio of bribery rates of traditional and modern firms increases to 1.7. These data patterns highlight the importance of bribery in explaining differences between countries in technology adoption,  productivity, and income along with existing explanations such as differences in entry and operating costs. In particular, the lower bribery rates of modern firms relative to traditional firms can potentially explain the higher adoption rates of more productive modern technology in high-income countries.

\subsection{The East Asian paradox}
\label{East_Asian_paradox}

In this section, we use the East Asian paradox to further illustrate our argument that the differences in bribery rates between modern and traditional firms can potentially explain the cross-country differences in technology adoption and economic performance. The term ``East Asian paradox" describes the failure of the predicted negative bribery-growth relationship for some East Asian countries such as China, Taiwan, South Korea, Japan and Indonesia (see \cite{Wedeman2003}). Historically, these East Asian countries have had levels of bribery on par with other Asian countries but much higher capital accumulation and growth rates. The literature has offered some explanations for this paradox, including the culture-specific ``relationship-based" nature of bribery in some of these countries (\cite{Vial2010}), the centralized nature of bribery in East Asia (\cite{Blackburn2009}), the large population size of these countries (\cite{RockBonnet2004}). We argue that differences in bribery rates in traditional and modern sectors within Asian countries can serve as an alternative explanation. 

In Table \ref{tab:bribery_asia}, we report bribery rates (in \% of annual sales) in traditional and modern sectors in selected Asian countries in 2012-2015 in our data\footnote{We do not report bribery data for Japan, South Korea, and Taiwan, because firms in these countries in our data do not report paying any bribes.}. In the table, China represents one of the outlying countries that have a high capital stock, a high growth rate, and a relatively high bribery rate. In fact, bribery rates across all countries in Table \ref{tab:bribery_asia} are similar in magnitude, but the ratios of bribery rates in the traditional and modern sectors differ. In China, bribery affects traditional firms much more heavily than modern firms, and the ratio of bribery rates in China is 1.8. This ratio is above the ratio of bribery rates in high-income countries (1.7) in our data. This contrasts with the much smaller ratio of bribery rates in the Philippines (1.0), which has had a lower growth rate than China despite having lower bribery rates in both sectors.  

\begin{table}[h]
\small
    \centering
    \caption{Bribery in traditional and modern sectors in selected countries in Asia in 2012-2015.}
    \label{tab:bribery_asia}
    \begin{tabular}{lcccccccc}
        \hline
        \hline
         {Country}& {Year} & {Traditional firms'} & {Modern firms'} & {$\frac{\tau_0}{\tau_1}$} & {Growth rate}\\
         & & {{bribe ($\tau_0$)} } & {bribe ($\tau_1$)} & & {in 2012-2015}\\
        \hline
{China} & {2012} & 5.2\% & 2.9\% & 1.8 & 7.5\%\\
{Pakistan} & {2013} & 6.1\% & 5.0\% & 1.2 & 4.2\%\\
{Philippines} & {2015} & 2.4\% & 2.4\% & 1.0 & 6.5\%\\
{Nepal} & {2013} & 5.3\% & 6.9\% & 0.8 & 4.3\%\\
{India} & {2014} & 2.4\% & 4.2\% & 0.6 & 6.5\%\\
        \hline
        \hline
    \end{tabular}
    \label{tab:asia_bribery}
\end{table}

The results in Table \ref{tab:bribery_asia} support the hypothesis that differences in bribery rates in modern and traditional sectors within individual countries can explain at least some cross-country variation in technology adoption, capital accumulation, and ultimately, growth. The relevant questions are about the exact mechanism through which bribery affects technology adoption in different countries and the magnitude of bribe effects on relevant economic indicators. We use counterfactual analysis to answer these questions. 

\section{Counterfactual analysis}
\label{counterfactual_analysis}

We focus on answering two questions. The first question asks whether bribery is harmful to economic performance. To answer this question, we study changes in steady-state values of the calibrated model in the counterfactual scenario, where bribery is completely eliminated for both modern and traditional firms. We address this question in section \ref{sec:no_bribery}. The second question asks whether bribery can serve as a potential source of higher economic performance. To answer this question, we study changes in steady-state values of the calibrated model in the counterfactual scenario, where bribery is eliminated only in the modern sector but is allowed to persist intact in the traditional sector. We address this question in section \ref{sec:no_modern_bribery}. In Section \ref{sec:no_traditional_bribery}, we demonstrate that the partial elimination of bribery can lead to substantial declines in capital stock and aggregate output relative to those when bribery is left intact. To illustrate this point, we study changes in steady-state values of the calibrated model in the counterfactual scenario, where bribery is eliminated only for traditional firms but is allowed to persist intact for modern firms. 

\subsection{No bribery in all sectors: the benchmark scenario} \label{sec:no_bribery}

In this section, we discuss the counterfactual case of complete elimination of bribery. In Table \ref{tab:no_bribe}, we report changes in percent of steady-state values of economic indicators of the calibrated model for countries grouped by income level. As in Table \ref{tab:summary_stats}, low-income countries are the countries with GDP per capita between 180.6 and 2,095.8 constant 2009 US dollars (lowest third in the data); middle-income countries are the countries with GDP per capita between 2,169 and 5,848 constant 2009 US dollars (middle third in the data); and high-income countries are the countries with GDP per capita between 5,936 and 72,735 constant 2009 US dollars (highest third in the data).

\begin{table}[h]
\small
    \centering
    \caption{Changes in steady-state values from the complete elimination of bribery.}
    \begin{tabular}{llccc}
        \hline
        \hline
        && {Low-income} & {Middle-income} & {High-income} \\
        && {countries} & {countries} & {countries} \\
        \hline
        {Aggregate}& {Intensive margin}  & 4.8\% & 2.2\% & 1.4\% \\
        {output} & {Extensive margin}  & 1.0\% & 0.4\% & -0.1\% \\
        &{Net change} & 5.8\% & 2.6\% & 1.3\% \\
        \hline
        {Aggregate} & {Intensive margin}  & 2.6\% & 1.2\% & 0.7\%\\
        {consumption} & {Extensive margin} & 0.4\% & 0.2\% & 0.0\%\\
        & {Net change} & 3.0\% & 1.4\% & 0.7\% \\
        \hline
        {Aggregate}& {Intensive margin}  & 7.9\% & 3.4\% & 2.0\% \\
        {capital} & {Extensive margin}  & 2.8\% & 1.1\% & -0.2\% \\
        & {Net change} & 10.7\% & 4.5\% & 1.8\% \\
        \hline
        {Average}&{Intensive margin}  & 6.8\% & 3.0\% & 2.1\%\\
        {wage} & {Extensive margin}  & 0.2\% & 0.1\% & 0.0\% \\
        & {Net change} & 7.0\% & 3.1\% & 2.1\% \\
        \hline
        {Modern firms'}&{Intensive margin}  & 0.6\%  & 0.2\%  & 0.0\% \\
        {share of output} & {Extensive margin}  & 1.6\%  & 0.7\%  & -0.2\% \\
        &{Net change} & 2.2\% & 0.9\% & -0.2\%\\
        \hline
        {Firms' entry}& & & & \\ 
        {(Number of firms)}&{Net change} & 5.6\% & 3.0\% & 2.0\% \\
        \hline
        {Fraction of}& & & &  \\
        {modern firms}&{Net change} & 1.9\% & 1.1\% & -0.2\% \\ 
        \hline
        \hline
    \end{tabular}
    \label{tab:no_bribe}
\end{table}

The signs of net changes in Table \ref{tab:no_bribe} indicate that the complete elimination of bribery increases aggregate production, consumption, capital stock, average wage, and encourages the entry of new firms in all groups of countries. The increases in steady-state values are largest in low-income countries, as in these countries firms pay the largest bribes and, therefore, benefit the most from the complete elimination of bribery. For example, in low-income countries, the capital stock increases by 10.7\%, the aggregate output by 5.8\%, and consumption by 3.0\%.  In high-income countries, where the bribe payments are smaller, the aggregate capital increases by only 1.8\%, the aggregate output by 1.3\%, and consumption by 0.7\%. These effects are consistent with findings in the literature that document the negative relationship between bribery rates, capital investment, and aggregate output (e.g. \cite{Mauro1995}, \cite{DeRosa2015}, \cite{Uberti2022}, \cite{FismanGurievIoramashviliPlekhanov2023}). Furthermore, the disproportionally large increases in capital stock, aggregate output, and consumption in low-income countries relative to middle- and high-income countries indicate that complete elimination of bribery can substantially reduce gaps in economic performance between poorer and wealthier countries.

Because in our model firms make entry, production and technological decisions, we decompose net changes from the complete elimination of bribery into intensive and extensive margins. Intensive margins measure the effects of the complete elimination of bribery on steady-state values arising from the higher entry of new firms and the higher utilization of inputs by incumbents, holding firms' choice of technology $j \in \{0,1\}$ the same as in the calibrated model. In Table \ref{tab:no_bribe}, the intensive margins are positive and represent the largest share of overall increases in steady-state values in all groups of countries. Extensive margins measure the effects of the complete elimination of bribery on steady-state values arising from the adoption of modern technology by entrants, holding entry rates and incumbents' use of inputs constant. In contrast to positive intensive margins, extensive margins vary from positive values in low- and middle-income countries to negative values in high-income countries. 

We obtain positive intensive margins in all groups of countries because the complete elimination of bribery encourages the production by incumbents and the entry of new firms by raising operating firms' revenue. As a result, the firms' aggregate demand for capital and labor increases, which is reflected in the positive intensive margins in entry rates, capital, and output. Further, since the aggregate supply of labor is fixed in our model, the increase in the firms' aggregate demand for labor unequivocally raises households' wages, which together with the larger aggregate output increases household consumption.      

We obtain negative extensive margins in high-income countries because in these countries modern firms pay smaller bribes than traditional firms, and the complete elimination of bribery makes the operation with modern technology relatively less profitable. As a result, in high-income countries, the adoption of modern technology by entrants and the share of aggregate output produced by modern firms decline by 0.2\%. Consequently, net increases in capital stock, aggregate output, and consumption in high-income countries are less than 2\%.

In low- and middle-income countries, modern firms pay higher bribes than traditional firms, and the complete elimination of bribery makes the operation with modern technology relatively more profitable. As a result, more entrants adopt modern technology, leading to positive extensive margins in capital stock, aggregate output, and the share of aggregate output produced by modern firms. Consequently, net increases in capital stock, output, and consumption in low- and middle-income countries are substantially greater than in high-income countries.       

The net increases in steady-state values from the complete elimination of bribery in Table \ref{tab:no_bribe} indicate that bribery is indeed harmful, but the impact of the intervention is not uniform in different groups of countries. In poorer countries, where bribery affects modern firms more heavily, the complete elimination of bribery raises incumbents' productivity, increases overall entry rates, and encourages the adoption of modern technology by entrants. In wealthier countries, where bribery affects traditional firms more heavily, the complete elimination of bribery raises incumbents' productivity, increases overall entry rates, but discourages the adoption of modern technology by entrants.

\subsection{No bribery in the modern sector: the high-output scenario} \label{sec:no_modern_bribery}

In this section, we discuss how elimination of bribery only for modern firms with bribery of traditional firms kept intact can encourage the adoption of modern technology, increase capital accumulation, and increase production beyond the values in the benchmark scenario without any bribery. In Table \ref{tab:no_modern_bribe}, we report changes in percent of steady-state values of economic indicators of the calibrated model for countries grouped by three income levels: low-income, middle-income, and high-income.

\begin{table}[h]
\small
    \centering
    \caption{Changes in steady-state values from the elimination of bribery only in the modern sector.}
    \begin{tabular}{llccc}
        \hline
        \hline
        && {Low-income} & {Middle-income} & {High-income} \\
        && {countries} & {countries} & {countries} \\
        \hline
        {Aggregate}& {Intensive margin} & 4.1\% & 1.9\% & 1.2\%\\
        {output} & {Extensive margin}  & 1.9\% & 0.9\% & 0.3\%\\
        &{Net change} & 6.0\% & 2.8\% & 1.5\% \\
        \hline
        {Aggregate} & {Intensive margin}  & 2.2\% & 1.1\% & 0.7\%\\
        {consumption} & {Extensive margin}  & 0.6\% & 0.4\% & 0.1\%\\
        & {Net change} & 2.8\% & 1.5\% & 0.8\% \\
        \hline
        {Aggregate}& {Intensive margin}  & 7.1\% & 3.1\% & 1.9\% \\
        {capital} & {Extensive margin}  & 5.6\% & 2.5\% & 0.7\% \\
        & {Net change} & 12.7\% & 5.6\% & 2.6\% \\
        \hline
        {Average}&{Intensive margin}  & 5.2\% & 2.3\% & 1.6\%\\
        {wage} & {Extensive margin}  & 0.3\% & 0.2\% & 0.0\%\\
        & {Net change} & 5.5\% & 2.5\% & 1.6\% \\
        \hline
        {Modern firms'}&{Intensive margin}  & 1.0\%  & 0.5\%  & 0.2\%  \\
        {share of output} & {Extensive margin}  & 3.4\%  & 1.6\%  & 0.5\% \\
        &{Net change} & 4.4\% & 2.1\% & 0.7\%\\                
        \hline
        {Firms' entry}& & & & \\
        {(Number of firms)} & {Net change} & 2.8\% & 2.0\% & 1.3\% \\
        \hline
        {Fraction of}& & & &\\
        {modern firms} & {Net change}  & 4.2\% & 2.6\% & 1.1\% \\     
        \hline
        \hline
    \end{tabular}
    \label{tab:no_modern_bribe}
\end{table}

As in the benchmark scenario, the elimination of bribery only for modern firms raises steady-state output, capital stock, consumption, entry rates, and adoption of modern technology across all groups of countries, and more so in low-income countries. Most importantly, the partial elimination of bribery in this scenario increases capital stock, adoption of modern technology, and aggregate output more than the complete elimination of bribery in all groups of countries. The net increase in the fraction of modern firms in this scenario exceeds that of the benchmark scenario by 2.3\% in low-income countries, by 1.5\% in middle-income countries, and by 1.3\% in high-income countries. Similarly, the net increase in capital stock in this scenario exceeds that of the benchmark scenario by 2.0\% in low-income countries, by 1.1\% in middle-income countries, and by 0.8\% in high-income countries. The net increase in aggregate output in this scenario exceeds that of the benchmark scenario by 0.2\% in all groups of countries.

The main source of the larger net increases in economic indicators in this scenario is substantially larger extensive margins. Extensive margins in aggregate capital, aggregate output, and the share of aggregate output produced by modern firms in Table \ref{tab:no_modern_bribe} are more than two times greater than those in Table \ref{tab:no_bribe} in all groups countries, while the intensive margins in the two tables are similar in size. This means that the main reason why capital stock and aggregate output increase more in this scenario is the higher adoption of modern technology by entrants rather than the higher entry of new firms and the more intensive utilization of inputs by incumbents. 

The larger increases in capital stock and output in this scenario do not translate to larger increases in consumption and wages. In fact, in low-income countries, the net increases in consumption and wages are smaller than under complete elimination of bribery. This is because in low-income countries, where the share of entrants adopting modern technology is the largest, the entrants' demand for capital input increases more than in the benchmark scenario, while the entrants' demand for labor input increases less than in the benchmark scenario. Consequently, wages and household consumption in low-income countries increase less in this scenario than in the benchmark scenario. As a result, the higher adoption of modern technology in low-income countries in this scenario leads to the allocation of a larger fraction of aggregate output to capital investment at the expense of reduced household consumption.

The results in Table \ref{tab:no_modern_bribe} indicate that partial elimination of bribery can lead to substantially larger increases in capital stock and output than complete elimination of bribery. When bribery is allowed to persist only in the traditional sector, its presence acts as a form of an incentive tax that redirects resources from the less productive traditional sector to the more productive modern sector in all groups of countries. In this way, the partial elimination of bribery serves as a source of higher capital accumulation and greater production, though at the expense of a smaller increase in consumption. 

\subsection{No bribery in the traditional sector: the low-output scenario} \label{sec:no_traditional_bribery}

In this section, we discuss the case of elimination of bribery only for traditional firms, holding the level of bribery of modern firms unchanged. We show that in contrast to the results in Section \ref{sec:no_modern_bribery}, the partial elimination of bribery in this scenario reduces aggregate output and capital stock not only below those of the benchmark scenario but also below those when bribery is allowed to persist intact. This case illustrates how differences in the bribery rates of modern and traditional firms can lead to a severe misallocation of resources within a single economy. We report changes in percent of steady-state values of the economic indicators of the calibrated model for low-, middle-, and high-income countries in Table \ref{tab:no_traditional_bribe}.

\begin{table}[h]
\small
    \centering
    \caption{Changes in steady-state values from the elimination of bribery only in the traditional sector.}
    \begin{tabular}{llccc}
        \hline
        \hline
        && {Low-income} & {Middle-income} & {High-income} \\
        && {countries} & {countries} & {countries} \\
        \hline
        {Aggregate}& {Intensive margin} & 0.7\% & 0.3\% & 0.2\% \\
        {output} & {Extensive margin}   & -0.9\% & -0.5\%  & -0.4\% \\
        &{Net change} & -0.2\% & -0.2\% & -0.2\% \\
        \hline
        {Aggregate} & {Intensive margin}  & 0.3\% & 0.1\% & 0.1\% \\
        {consumption} & {Extensive margin} & -0.3\% & -0.2\%  & -0.1\% \\
        & {Net change} & 0.0\% & -0.1\% & 0.0\% \\
        \hline
        {Aggregate}& {Intensive margin}  & 0.7\% & 0.3\% & 0.1\% \\
        {capital} & {Extensive margin}  & -2.4\% & -1.4\%  & -0.9\%  \\
        & {Net change} & -1.7\% & -1.1\% & -0.8\% \\
        \hline
        {Average}&{Intensive margin}  & 1.7\% & 0.7\% & 0.4\% \\
        {wage} & {Extensive margin}  & 0.1\% & 0.1\%  & 0.1\% \\
        & {Net change} & 1.8\% & 0.8\% & 0.5\% \\
        \hline
        {Modern firms'}&{Intensive margin}  & -0.4\%  & -0.2\%  & -0.2\% \\
        {share of output} & {Extensive margin}  & -2.0\%  & -1.2\%  & -0.8\% \\
        &{Net change} & -2.4\% & -1.4\% & -1.0\%\\        
        \hline
        {Firms' entry}& & & & \\
        {(Number of firms)} &{Net change} & 3.0\% & 1.1\% & 0.8\% \\
        \hline
        {Fraction of}& & & & \\
        {modern firms} & {Net change}  & -2.1\% & -1.7\% & -1.4\% \\     
        \hline
        \hline
    \end{tabular}
    \label{tab:no_traditional_bribe}
\end{table}

In contrast to results in the benchmark scenario (Table \ref{tab:no_bribe}) and the high-output scenario (Table \ref{tab:no_modern_bribe}), the elimination of bribery only for traditional firms reduces the steady-state capital stock by 1.7\% in low-income countries, by 1.1\% in middle-income countries, and by 0.8\% in high-income countries, while at the same time reducing output by 0.2\% in all groups of countries. These results show that the partial elimination of bribery, while reducing the overall burden of bribery, can lead to significant declines in capital stock and aggregate output. 

The main source of these declines are large negative extensive margins due to fewer entrants adopting modern technology. In contrast to the benchmark scenario and the high-output scenario, the elimination of bribery only for traditional firms makes operation of firms with modern technology relatively less profitable, leading to more incumbents adopting traditional technology. This increases the demand for labor input and reduces the demand for capital input. These effects are reflected in negative extensive margins in the share of output produced by modern firms in Table \ref{tab:no_traditional_bribe}. The higher demand for labor input is reflected in positive intensive and extensive margins in wages. 

Furthermore, the elimination of bribery only for traditional firms increases the net entry rates of firms (number of operating firms) by 3.0\% in low-income countries, 1.1\% in middle-income countries, and 0.8\% in high-income countries. These increases are driven by the increased adoption of the cost-free traditional technology by entrants. 

In contrast to the results in Sections \ref{sec:no_bribery} and \ref{sec:no_modern_bribery}, the partial elimination of bribery in this scenario leads to declines in capital accumulation, output, and almost no change in consumption relative to the scenario without any bribery. As in the high-output scenario of Section \ref{sec:no_modern_bribery}, bribery acts as a form of an incentive tax. But, since bribery affects only modern firms in this scenario, its presence diverts resources away from the more productive modern sector toward the less productive traditional sector mainly by discouraging adoption of the more productive modern technology by entrants. As a result, the partial elimination of bribery in this scenario leads to a large misallocation of resources manifested by substantial declines in capital stock in all groups of countries. Most importantly, the declines in capital stock are more pronounced in poorer countries than in wealthier countries, which means that the elimination of bribery only for traditional firms, in fact, expands cross-country gaps in capital stock and output.   

\section{Conclusion}
\label{conclusion}

We study how bribery, differentiated by types of firms, impacts technology adoption, capital accumulation, consumption, and output across countries. We document large differences in bribery rates faced by modern and traditional firms in different countries and show that these differences can explain at least some variation in economic performance between poorer and wealthier countries. 

We construct a dynamic stochastic general equilibrium model with differentiated technology-specific bribery rates and heterogeneous firms that endogenously choose between modern and traditional production technologies. Consistently with existing results in the literature, we find that bribery indeed stifles production by discouraging entry of new firms and by suppressing utilization of inputs by incumbent firms. However, when bribery affects traditional firms more heavily than modern firms, the presence of bribery encourages the adoption of modern technology by entrants, thereby raising capital accumulation and aggregate output beyond those of the benchmark case without bribery. Furthermore, the disproportionately large increases in capital stock and aggregate output in low-income countries relative to high-income countries indicate that anti-bribery policies that prioritize modern sectors of the economy can not only substantially increase adoption of modern technology, capital accumulation and aggregate output, but also narrow cross-country gaps in capital stock and aggregate output.

When bribery affects modern firms more heavily than traditional firms, the presence of bribery encourages the adoption of labor-intensive traditional technology by entrants, thereby discouraging capital accumulation and reducing aggregate output. As a result, an intervention leading to the elimination of bribery only for traditional firms may, in fact, lead to declines in capital accumulation, consumption, aggregate output, and widen cross-country gaps between wealthier and poorer countries.


In this paper, we do not make explicit assumptions about household utility to minimize the number of assumptions underlying our main findings. Our approach limits our ability to discuss the welfare implications, particularly the changes in the household welfare due to partial and complete elimination of bribery. Furthermore, since our analysis focuses exclusively on steady-state outcomes, we do not consider the growth effects and the transition dynamics associated with policy interventions. We leave these important welfare and dynamic aspects for future research.

The results in this paper prescribe how anti-corruption campaigns should be designed and implemented to increase capital accumulation, household consumption, and aggregate production. In particular, if the goal of an anti-corruption campaign is to increase aggregate production through adoption of more productive technologies or if the implementation of anti-bribery policies is limited by budget constraints, then the reduction in bribery and corruption in high-productivity capital-intensive sectors of the economy should be given the priority. 

\newpage

\bibliographystyle{aer}
\bibliography{acompat,bibliography_jafar}

\newpage

\section{Appendix}
\subsection{Additional quantitative analysis}
In this section, we discuss additional counterfactual exercises that do not directly involve adjustments in bribery rates. We do so to identify policies that can help narrow the gaps in aggregate output and capital stock between low-income and high-income countries in the presence of bribery. 

\subsubsection{The increase in the productivity of modern firms}
\label{A1_increase}
We explore how an increase in the productivity of modern firms ($A_1$), while keeping bribery intact in both sectors, affects the performance of firms in countries in different income groups. This intervention helps to answer the question of whether freely disseminated technological innovation can narrow the gap between low- and high-income countries in the presence of bribery. Specifically, we ask whether the advancement of the generative AI technology, assuming that all firms in all countries have free access to it, could help low-income countries reach income levels of high-income countries, if bribery rates in all countries remained the same as in the calibrated model.      

We consider the effect of the uniform 20\% increase in the productivity of modern firms ($A_1$) in all countries and report changes in steady-state values in Table \ref{tab:A1_20p_increase}. The results in Table \ref{tab:A1_20p_increase} indicate that the 20\% increase in the productivity of modern firms in all countries substantially increases the entry rates of firms, the adoption of modern technology by entrants, the capital stock, average wages, aggregate consumption, and aggregate output in all groups of countries.

\begin{table}[h]
\small
    \centering
    \caption{Changes in steady-state values from the increase in $A_1$ by 20\%.}
    \begin{tabular}{llccc}
        \hline
        \hline
        && {Low-income} & {Middle-income} & {High-income} \\
        && {countries} & {countries} & {countries} \\
        \hline
        {Aggregate}& {Intensive margin} & 15.0\% & 17.0\% & 19.8\%\\
        {output} & {Extensive margin}  & 5.8\% & 5.3\% & 2.9\%\\
        &{Net change} & 20.8\% & 22.3\% & 22.7\% \\
        \hline
        {Aggregate} & {Intensive margin}  & 14.2\% & 16.3\% & 19.4\%\\
        {consumption} & {Extensive margin} & 2.7\% & 2.6\%  & 1.4\%\\
        & {Net change} & 16.9\% & 18.9\% & 20.8\% \\
        \hline
        {Aggregate}& {Intensive margin}  & 18.3\% & 19.7\% & 21.2\% \\
        {capital} & {Extensive margin}  & 14.3\% & 12.2\%  & 6.1\% \\
        & {Net change} & 32.6\% & 31.9\% & 27.3\% \\
        \hline
        {Average}&{Intensive margin}  & 13.1\% & 15.3\% & 18.9\%\\
        {wage} & {Extensive margin}  & 1.1\% & 1.1\%  & 0.6\%\\
        & {Net change} & 14.2\% & 16.4\% & 19.5\% \\
        \hline
        {Modern firms'}&{Intensive margin}  & 3.6\%  & 3.0\%  & 1.7\%\\
        {share of output} & {Extensive margin}  & 8.4\%  & 7.2\%  & 3.6\%\\
        &{Net change} & 12.0\% & 10.2\% & 5.3\%\\        
        \hline
        {Firms' entry} & & & & \\
        {(Number of firms)} & {Net change} & 7.7\% & 12.2\% & 16.8\% \\
        \hline
        {Fraction of}& & & &\\
        {modern firms} & {Net change}  & 12.3\% & 13.3\% & 8.8\% \\    
        \hline
        \hline
    \end{tabular}
    \label{tab:A1_20p_increase}
\end{table}

 The highest increases in capital accumulation and adoption of modern technology from the intervention are in low-income countries (32.6\% and 12.3\%, respectively), and the smallest increases in capital accumulation and adoption of modern technology are in high-income countries (27.3\% and 8.8\%, respectively). However, the highest increases in aggregate output and consumption are in high-income countries (22.7\% and 20.8\%, respectively), and the lowest increases in aggregate output and consumption are in low-income countries (20.8\% and 16.9\%, respectively). 
 
 The main reason for the larger increases in aggregate output and consumption in high-income countries relative to those in low-income countries is the higher intensive margins in aggregate output and consumption in high-income countries. In particular, the higher intensive margin in aggregate output in high-income countries (19.8\%) than in low-income countries (15.0\%) indicates that the 20\% increase in $A_1$ helps incumbents in high-income countries more than in low-income countries. This is because more incumbents in high-income countries operate with modern technology and, therefore, more firms in high-income countries benefit from the increase in $A_1$. 
 
 The results in Table \ref{tab:A1_20p_increase} indicate that in the presence of bribery, the dissemination of technological innovation (e.g. generative AI), even if frictionless and costless, does not necessarily lead to the convergence of consumption and incomes across countries with different levels of wealth. In fact, our results indicate that while freely disseminated technological innovation narrows gaps in capital accumulation and adoption of modern technology between low- and high-income countries, it expands gaps in aggregate output and consumption and makes the low-income countries fall even further behind the middle- and high-income countries in aggregate output.
 
\subsubsection{The decrease in the firms' cost of entry}

We explore how a decrease in the firms' cost of entry ($c_e$), while keeping bribery in both sectors intact, affects capital accumulation, consumption, and aggregate output in countries in different income groups. We consider the effect of the uniform 20\% decrease in the firms' cost of entry. The resulting changes in steady-state values are reported in Table \ref{tab:ce_20p_decrease}. 

As with the increase in $A_1$, the decrease in $c_e$ encourages the entry of new firms, increases capital stock, aggregate consumption, and aggregate output across countries in all income groups. However, the negative extensive margins in aggregate capital and aggregate output indicate that these increases arise from higher utilization of inputs by incumbent firms, rather than from the adoption of modern technology by entrant firms. In fact, the fraction of modern firms and their share of output decrease in all countries, as $c_e$ is reduced.    

Most importantly, the uniform 20\% reduction in entry cost $c_e$ expands the gaps in capital stock, aggregate consumption, and aggregate income between poorer and wealthier countries. Specifically, in low-income countries, the capital stock, aggregate consumption and production increase by 10.0\%, 13.9\%, and 13.0\%, respectively, while in high-income countries, the capital stock, aggregate consumption and output increase by 13.5\%, 14.4\% and 14.5\%, respectively. Hence, while the reduction in $c_e$ stimulates economic activity in countries in all income groups, these increases are driven mainly by the higher entry of firms into traditional sectors. Since wealthier countries already have larger fractions of modern firms, and the decline in $c_e$ mainly encourages entrants to adopt the less productive traditional technology, the gaps in aggregate output and capital stock between poorer and wealthier countries expand. In this regard, the effect of the reduction in $c_e$ is closest to the effect of the elimination of bribery in the traditional sector in Section \ref{sec:no_traditional_bribery}, where the gap in aggregate capital between low-income and high-income countries also widens. 

\begin{table}[h]
\small
    \centering
    \caption{Changes in steady-state values from the decrease in $c_e$ by 20\%.}
    \begin{tabular}{llccc}
        \hline
        \hline
        && {Low-income} & {Middle-income} & {High-income} \\
        && {countries} & {countries} & {countries} \\
        \hline
        {Aggregate}& {Intensive margin} & 14.9\% & 15.7\% & 15.3\%\\
        {output} & {Extensive margin}  & -1.9\% & -1.4\% & -0.8\%\\
        &{Net change} & 13.0\% & 14.3\% & 14.5\%\\
        \hline
        {Aggregate} & {Intensive margin}  & 13.9\% & 14.6\% & 14.3\%\\
        {consumption} & {Extensive margin} & 0.0\% & 0.0\% & 0.1\%\\
        & {Net change} & 13.9\% & 14.6\% & 14.4\%\\
        \hline
        {Aggregate}& {Intensive margin}  & 15.6\% & 16.3\% & 15.6\%\\
        {capital} & {Extensive margin}  & -5.6\% & -3.9\% & -2.1\%\\
        & {Net change} & 10.0\% & 12.4\% & 13.5\%\\
        \hline
        {Average}&{Intensive margin}  & 14.4\% & 15.3\% & 15.1\%\\
        {wage} & {Extensive margin}  & 0.5\% & 0.3\% & 0.2\%\\
        & {Net change} & 14.9\% & 15.6\% & 15.3\%\\
        \hline
        {Modern firms'}&{Intensive margin}  & 0.8\%  & 0.7\%  & 0.4\%\\
        {share of output} & {Extensive margin}  & -4.4\%  & -3.1\%  & -1.8\%\\
        &{Net change} & -3.6\% & -2.4\% & -1.4\%\\        
        \hline
        {Firms' entry}& & & &\\
        {(Number of firms)} & {Net change}  & 43.6\% & 41.1\% & 38.1\%\\
        \hline
        {Fraction of}& & & &\\
        {modern firms} & {Net change}  & -5.9\% & -6.2\% & -6.0\% \\    
        \hline
        \hline
    \end{tabular}
    \label{tab:ce_20p_decrease}
\end{table}

\subsubsection{The decrease in the operating cost of modern technology}
 
We explore how a decrease in the operating cost of modern technology ($c_1$), while keeping the bribery in both sectors intact, affects capital accumulation and aggregate output in countries in different income groups. We consider the effect of the uniform 20\% decrease in the operating cost of modern technology and report the resulting changes in steady-state values in Table \ref{tab:c1_20p_decrease}. 

\begin{table}[h]
\small
    \centering
    \caption{Changes in steady-state values from the decrease in $c_1$ by 20\%.}
    \begin{tabular}{llccc}
        \hline
        \hline
        && {Low-income} & {Middle-income} & {High-income} \\
        && {countries} & {countries} & {countries} \\
        \hline
        {Aggregate}& {Intensive margin} & 7.8\% & 7.7\% & 8.8\%\\
        {output} & {Extensive margin}  & 4.3\% & 3.3\% & 1.9\%\\
        &{Net change} & 12.1\% & 11.0\% & 10.7\% \\
        \hline
        {Aggregate} & {Intensive margin}  & 8.5\% & 8.4\% & 9.6\%\\
        {consumption} & {Extensive margin} & 1.8\% & 1.4\% & 0.8\%\\
        & {Net change} & 10.3\% & 9.8\% & 10.4\%\\
        \hline
        {Aggregate}& {Intensive margin}  & 8.1\% & 7.9\% & 8.9\%\\
        {capital} & {Extensive margin}  & 10.1\% & 7.5\%  & 4.0\%\\
        & {Net change} & 18.2\% & 15.4\% & 12.9\% \\
        \hline
        {Average}&{Intensive margin}  & 7.6\% & 7.5\% & 8.7\%\\
        {wage} & {Extensive margin}  & 0.8\% & 0.6\% & 0.4\%\\
        & {Net change} & 8.5\% & 8.1\% & 9.1\%\\
        \hline
        {Modern firms'}&{Intensive margin}  & 0.4\%  & 0.3\%  & 0.2\%\\
        {share of output} & {Extensive margin}  & 6.6\%  & 5.1\%  & 2.7\%\\
        &{Net change} & 7.0\% & 5.4\% & 2.9\%\\        
        \hline
        {Firms' entry}& & & &\\
        {(Number of firms)} & {Net change} & 7.7\% & 10.6\% & 14.1\%\\
        \hline
        {Fraction of}& & & &\\
        {modern firms} & {Net change}  & 10.5\% & 10.2\% & 8.0\% \\
        \hline
        \hline
    \end{tabular}
    \label{tab:c1_20p_decrease}
\end{table}

The results in Table \ref{tab:c1_20p_decrease} are similar to those in Table \ref{tab:A1_20p_increase}, as both the decrease in $c_1$ and the increase in $A_1$ make operation with the modern technology more appealing to entrant firms. The reduction of 20\% in $c_1$ increases the overall entry rates of firms, the adoption of modern technology by entrants, the capital stock, average wages, aggregate consumption and aggregate output in all groups of countries. However, in contrast to the results in Table \ref{tab:A1_20p_increase}, the results in Table \ref{tab:c1_20p_decrease} indicate that the 20\% decrease in $c_1$ raises aggregate output in low-income countries (12.1\%) more than in middle-income (11.0\%) and high-income (10.7\%) countries.

The reason for this difference are the intensive margins in the aggregate output. Note that unlike the 20\% increase in $A_1$, the 20\% decline in $c_1$ yields similar intensive margins in aggregate output in low-income (7.8\%), middle-income (7.7\%), and high-income (8.8\%) countries. Coupled with the much higher extensive margins in low-income countries, the net effect of the intervention on aggregate output turns out to be much larger in low-income countries than in middle- and high-income countries.


The results in Table \ref{tab:c1_20p_decrease}  indicate that in the presence of bribery, the reduction in the operating cost of modern technology not only expands the capital stock, aggregate consumption and aggregate output in countries across all income groups, but also narrows the gaps in capital stock and aggregate output between the poorer and wealthier countries. Furthermore, the comparison of results in Tables \ref{tab:c1_20p_decrease}, \ref{tab:A1_20p_increase}, and \ref{tab:ce_20p_decrease} demonstrates that while all three policy interventions (increase in $A_1$, reduction in $c_e$, and reduction in $c_1$) stimulate economic activity, only reduction in $c_1$ narrows gaps in consumption, capital stock, and output between wealthier and poorer countries. In this respect, the policies that reduce operating costs of modern technology are closest to the growth-enhancing policies that aim at reducing bribery in the modern sector (see Section \ref{sec:no_modern_bribery}). 
Consequently, policies that stimulate the adoption of modern technology (reduction in $A_1$) without simultaneously stimulating the operation with modern technologies (reduction in $c_1$) may not be sufficient to narrow the gaps in output between poorer and wealthier countries.   

\subsubsection{The decrease in the uncertainty of bribe demands} \label{sec:uncertain_bribery}

We explore how a decline in the uncertainty of bribe demands, while keeping the expected bribery rate constant, affects the performance of firms in low-income countries. The literature has argued that bribery is more distortionary than taxation, precisely because bribery is more unpredictable than taxation (see \cite{Wei1997}, \cite{Campos1999}, \cite{FismanSvensson2007}). We test this argument in the context of our model by calculating the decreases in steady-state values caused by the introduction of bribery with different levels of uncertainty, while holding the expected bribery rate in the traditional and modern sectors constant at 5\%. 

We report results in Table \ref{tab:variance_bribery_decrease}. The reported numbers are calculated as differences between variables of interest, when there is no bribery in both sectors and when firms in both modern and traditional sectors face bribes of size $\tau$ (in percent of total sales) with probability $P$. 

Note that as the variance in bribe requests (measured by $P$) declines, the capital stock, the firms' entry rate, the fraction of modern firms, and their share of output decline at a lower rate. For example, aggregate production declines by 11.4\%, when bribes of 10\% of total sales are demanded with probability 50\%, and by 8.6\%, when bribes of 50\% of total sales are demanded with probability 10\%. The similar pattern holds for declines in capital stock. The more uncertain but smaller bribe request ($P=0.5$, $\tau=10$) reduces the capital stock by 19.3\% from the benchmark value when there is no bribery, while the more certain but larger bribe request ($P=0.1$, $\tau=50$) reduces the capital stock only by 11.5\%. Similar patterns hold for declines in entry rates, wages, and fractions of modern firms. These results are consistent with previous findings in the literature that emphasize that uncertainty in bribe demands is one of the main sources of distortions caused by bribery (e.g. \cite{Wei1997}, \cite{Campos1999}) and in the literature that studies the effect of informational frictions on firm-level misallocation (e.g. \cite{DavidHopenhaynVenkateswaran2016}, \cite{DavidVenkateswaran2019}). 

\begin{table}[h]
\small
    \begin{center}
    \caption{Declines in steady-state values from the introduction of bribery in low-income countries.}
    \label{tab:variance_bribery_decrease}
    \begin{tabular}{lccc}
        \hline
        \hline
         & {$P=0.5$}, & {$P=0.25$} & {$P=0.1$} \\
         & {\(\tau\)=10} & {\(\tau\)=20} & {\(\tau\)=50} \\
        \hline
        {Aggregate output} & -11.4\% & -10.7\% & -8.6\% \\
        {Aggregate consumption}& -5.6\% & -6.0\% & -6.5\% \\
        {Aggregate capital} & -19.3\% & -17.0\% & -11.5\% \\
        {Average wage} & -14.9\% & -13.3\% & -9.4\% \\
        {Modern firms' share of output} & -3.1\% & -2.7\% & -1.7\% \\
        {Firms' entry (Number of firms)} & -11.2\% & -10.2\% & -7.5\% \\
        {Fraction of modern firms} & -2.5\% & -2.1\% & -1.3\% \\
        \hline
        \hline
    \end{tabular}
    \end{center}
    \textbf{Notes:} The table reports declines in steady-state values relatively to the case without bribery when bribes of size $\tau$ are demanded with probability $P$ in low-income countries. The values in the first column ($P=0.5$, $\tau=10$) correspond to declines when bribes of 10\% of total sales are demanded in both sectors with probability $P=0.5$. The values in the second column ($P=0.25$, $\tau=20$) correspond to declines when bribes of 20\% of total sales are demanded in both sectors with probability $P=0.25$. The values in the third column ($P=0.1$, $\tau=50$) correspond to declines when bribes of 50\% of total sales are demanded in both sectors with probability $P=0.1$. All three scenarios have expected bribes fixed at 5\%.  
\end{table}

\newpage
\section{Data appendix: Not for publication}

In this section, we describe the data that we use to calibrate the model. We use data from the World Bank Enterprise Surveys (WBES)\footnote{See \url{www.enterprisesurveys.org} for a description of the methodology of data collection.}. The data contain survey responses from firms' managers operating in 148 countries between 2006 and 2023. Because our data contain firm-level surveys from the same countries in multiple years, we treat each survey from each country in a specific year as a survey from a unique country. In particular, we do not merge surveys coming from same countries in different years. We do so because the data do not contain unique identifiers for firms, and the same firms could be represented in surveys from different years. 

In total, we have 276 surveys. The data from all surveys were converted to US dollars and deflated to constant prices from 2009. The summary statistics are reported in Table \ref{tab:appendix_data_summary}. The table reports the code for each country in the first column, the year when the survey was conducted in the second column, and the size of each survey (the number of surveyed firms) in the third column. 

We follow \cite{Farrokhi2024} and employ the k-means clustering algorithm to classify the firms in each survey as modern and traditional firms. Given the distribution of firms' value of capital per worker (capital-labor ratio) in each survey, firms with the value of capital per worker above a certain threshold are classified as modern and below the threshold as traditional. The survey-specific threshold $\omega^*$ minimizes the sum of variances of capital-labor ratios of traditional ($\kappa_0(\omega)$) and modern ($\kappa_1(\omega)$) firms in each survey: 

\begin{equation}
    \omega^* = \arg \min \sum_{\omega\in\Omega_0} (\kappa_0(\omega) - \bar \kappa_0)^2 +\sum_{\omega\in\Omega_1} (\kappa_1(\omega) - \bar \kappa_1)^2
\end{equation} 
where \(\bar \kappa_j\) is the average capital-labor ratio of firms with technology $j$ in the survey, \(\Omega_0 = \{\omega|\kappa(\omega) <\kappa(\omega^*)\}\) is the set of traditional firms in the survey, and \(\Omega_1 = \{\omega|\kappa(\omega) >\kappa(\omega^*)\}\) is the set of modern firms in the survey. We report estimated shares of modern firms in each survey in the fourth column in Table \ref{tab:appendix_data_summary}.

Having classified the firms in each survey into traditional and modern, we calculate average bribes and frequencies of bribe requests for each groups of firms in each survey. In Table \ref{tab:appendix_data_summary}, we report calibrated bribes and frequencies of bribe requests for traditional firms in the fifth and sixth columns, respectively. Similarly, we report calibrated bribes and frequencies of bribe requests for modern firms in the seventh and eighths columns, respectively. 

We drop all surveys with less than 10 firms and surveys, where traditional and modern firms do not report bribes. Next, we drop observations with missing values and observations with outliers in the remaining surveys. Observations with outliers were identified based on the three-standard-deviation rule applied to reported sales, capital, cost of labor, and cost of intermediate inputs. This leaves us with 42,447 firms from 148 unique countries. Lastly, we drop surveys with negative calibrated operating cost in the modern sector ($c1$). We are left with data from 36,619 firms from 219 surveys from 124 unique countries. We use these data in the calibration exercise.

\begin{table}[]
\caption{Summary statistics (all surveys) in 2006-2023.}
\label{tab:appendix_data_summary}
\begin{tabular}{llllllll}
\hline
\hline
ISO & Year & \begin{tabular}[c]{@{}l@{}}Number \\ of firms\end{tabular} & \begin{tabular}[c]{@{}l@{}}Share of \\ modern \\firms\end{tabular} & \begin{tabular}[c]{@{}l@{}}Bribe in \\ traditional \\ sector\end{tabular} & \begin{tabular}[c]{@{}l@{}}Probability of \\ bribe request in \\ traditional sector\end{tabular} & \begin{tabular}[c]{@{}l@{}}Bribe in \\ modern\\ sector\end{tabular} & \begin{tabular}[c]{@{}l@{}}Probability of \\ bribe request in \\ modern sector\end{tabular} \\
\hline
AFG & 2008 & 35 & 22.9\% & 6.9\% & 29.6\% & 2.3\% & 37.5\% \\
ALB & 2019 & 37 & 38.6\% & 9.8\% & 44.2\% & 6.8\% & 32.7\% \\
AGO & 2006 & 172 & 36.6\% & 7.4\% & 46.9\% & 6.9\% & 50.2\% \\
AGO & 2010 & 10 & 38.6\% & 1.0\% & 18.0\% & 0.0\% & 0.0\% \\
ATG & 2010 & 20 & 89.7\% & 0.0\% & 0.0\% & 0.0\% & 0.0\% \\
ARG & 2006 & 190 & 82.3\% & 5.8\% & 28.0\% & 8.9\% & 27.7\% \\
ARG & 2010 & 334 & 76.8\% & 3.2\% & 25.8\% & 7.4\% & 7.2\% \\
ARG & 2017 & 126 & 85.9\% & 0.0\% & 0.0\% & 1.1\% & 13.8\% \\
ARM & 2013 & 14 & 76.9\% & 0.0\% & 0.0\% & 1.5\% & 16.3\% \\
ARM & 2020 & 87 & 64.3\% & 0.0\% & 0.0\% & 0.0\% & 0.0\% \\
AUT & 2021 & 140 & 100.0\% & 0.0\% & 0.0\% & 0.0\% & 0.0\% \\
BHS & 2010 & 25 & 90.7\% & 0.0\% & 0.0\% & 1.8\% & 8.2\% \\
BGD & 2022 & 288 & 22.0\% & 10.4\% & 2.6\% & 4.7\% & 5.4\% \\
BRB & 2010 & 37 & 89.8\% & 0.0\% & 0.0\% & 3.0\% & 1.6\% \\
BRB & 2023 & 28 & 85.9\% & 0.0\% & 0.0\% & 0.0\% & 0.0\% \\
BLR & 2013 & 39 & 36.0\% & 5.7\% & 7.5\% & 10.0\% & 8.9\% \\
BLR & 2018 & 199 & 61.4\% & 0.0\% & 0.0\% & 0.0\% & 0.0\% \\
BEL & 2020 & 174 & 99.0\% & 0.0\% & 0.0\% & 2.0\% & 0.2\% \\
BLZ & 2010 & 54 & 77.1\% & 0.0\% & 0.0\% & 3.0\% & 1.5\% \\
BEN & 2016 & 33 & 32.0\% & 12.9\% & 22.3\% & 7.7\% & 50.5\% \\
BTN & 2015 & 47 & 36.5\% & 25.0\% & 1.4\% & 0.0\% & 0.0\% \\
BOL & 2006 & 132 & 51.6\% & 8.6\% & 34.5\% & 9.2\% & 54.7\% \\
BOL & 2010 & 23 & 64.3\% & 20.0\% & 1.8\% & 1.0\% & 73.1\% \\
BOL & 2017 & 34 & 52.9\% & 10.0\% & 4.9\% & 4.8\% & 3.7\% \\
BIH & 2013 & 48 & 97.6\% & 0.0\% & 0.0\% & 1.4\% & 11.0\% \\
BIH & 2019 & 27 & 87.9\% & 0.0\% & 0.0\% & 5.0\% & 9.1\% \\
BIH & 2023 & 40 & 96.5\% & 0.0\% & 0.0\% & 0.0\% & 0.0\% \\
BWA & 2006 & 80 & 73.7\% & 3.9\% & 20.5\% & 11.3\% & 20.3\% \\
BWA & 2010 & 37 & 74.7\% & 0.0\% & 0.0\% & 11.1\% & 5.7\% \\
BWA & 2023 & 97 & 49.7\% & 22.0\% & 8.1\% & 22.4\% & 17.5\% \\
BRA & 2009 & 89 & 70.4\% & 8.2\% & 100.0\% & 2.9\% & 100.0\% \\
BGR & 2007 & 270 & 64.3\% & 5.8\% & 8.9\% & 4.1\% & 14.2\% \\
BGR & 2013 & 38 & 47.2\% & 0.0\% & 0.0\% & 5.0\% & 5.1\% \\
BGR & 2019 & 105 & 55.4\% & 10.0\% & 0.7\% & 6.2\% & 8.7\% \\
BGR & 2023 & 157 & 81.0\% & 20.0\% & 1.1\% & 2.9\% & 2.4\% \\
BFA & 2009 & 16 & 55.9\% & 10.0\% & 14.8\% & 40.0\% & 28.8\% \\
BDI & 2006 & 84 & 20.9\% & 8.0\% & 49.7\% & 9.7\% & 32.4\% \\
BDI & 2014 & 34 & 53.2\% & 2.4\% & 9.5\% & 2.7\% & 52.4\% \\
KHM & 2016 & 50 & 26.4\% & 4.0\% & 78.9\% & 1.1\% & 96.7\% \\
KHM & 2023 & 147 & 29.4\% & 3.2\% & 89.5\% & 3.8\% & 16.3\% \\
CMR & 2009 & 41 & 40.5\% & 8.6\% & 47.4\% & 10.0\% & 49.9\% \\
CMR & 2016 & 43 & 49.8\% & 5.0\% & 33.7\% & 11.7\% & 72.5\% \\
TCD & 2018 & 41 & 6.6\% & 14.3\% & 38.1\% & 11.9\% & 90.7\% \\
TCD & 2023 & 55 & 40.0\% & 28.0\% & 17.5\% & 16.2\% & 18.6\% \\
\hline
\hline
\end{tabular}
\end{table}

\begin{table}[]
\begin{tabular}{llllllll}
\hline
\hline
ISO & Year & \begin{tabular}[c]{@{}l@{}}Number \\ of firms\end{tabular} & \begin{tabular}[c]{@{}l@{}}Share of \\ modern \\firms\end{tabular} & \begin{tabular}[c]{@{}l@{}}Bribe in \\ traditional \\ sector\end{tabular} & \begin{tabular}[c]{@{}l@{}}Probability of \\ bribe request in \\ traditional sector\end{tabular} & \begin{tabular}[c]{@{}l@{}}Bribe in \\ modern\\ sector\end{tabular} & \begin{tabular}[c]{@{}l@{}}Probability of \\ bribe request in \\ modern sector\end{tabular} \\
\hline
CHL & 2006 & 343 & 77.0\% & 5.1\% & 3.9\% & 13.9\% & 6.2\% \\
CHL & 2010 & 539 & 70.4\% & 0.0\% & 0.0\% & 2.9\% & 2.5\% \\
CHN & 2012 & 899 & 69.7\% & 5.2\% & 3.3\% & 2.9\% & 6.8\% \\
COL & 2006 & 352 & 51.6\% & 13.3\% & 9.1\% & 14.4\% & 3.8\% \\
COL & 2010 & 457 & 62.6\% & 8.4\% & 1.0\% & 13.8\% & 6.0\% \\
COL & 2017 & 304 & 69.6\% & 9.6\% & 11.5\% & 7.9\% & 3.4\% \\
COL & 2023 & 124 & 66.2\% & 9.5\% & 16.2\% & 2.2\% & 2.5\% \\
CRI & 2010 & 21 & 60.1\% & 1.0\% & 17.6\% & 13.4\% & 30.0\% \\
CRI & 2023 & 37 & 72.7\% & 0.0\% & 0.0\% & 0.0\% & 0.0\% \\
HRV & 2007 & 132 & 93.3\% & 1.0\% & 6.3\% & 11.0\% & 4.0\% \\
HRV & 2013 & 65 & 87.9\% & 0.0\% & 0.0\% & 9.2\% & 9.3\% \\
HRV & 2019 & 96 & 95.3\% & 0.0\% & 0.0\% & 1.0\% & 6.4\% \\
HRV & 2023 & 69 & 86.4\% & 0.0\% & 0.0\% & 0.0\% & 0.0\% \\
CYP & 2019 & 31 & 87.8\% & 0.0\% & 0.0\% & 1.0\% & 3.1\% \\
DNK & 2020 & 473 & 96.7\% & 0.0\% & 0.0\% & 2.0\% & 1.1\% \\
DMA & 2010 & 13 & 85.2\% & 0.0\% & 0.0\% & 0.0\% & 0.0\% \\
DOM & 2010 & 65 & 80.7\% & 0.0\% & 0.0\% & 0.0\% & 0.0\% \\
DOM & 2016 & 21 & 47.8\% & 50.0\% & 26.3\% & 0.0\% & 0.0\% \\
COD & 2006 & 104 & 14.1\% & 5.6\% & 83.5\% & 7.9\% & 86.6\% \\
COD & 2010 & 47 & 32.6\% & 8.2\% & 63.2\% & 7.1\% & 67.6\% \\
COD & 2013 & 74 & 11.7\% & 11.0\% & 48.1\% & 9.6\% & 67.9\% \\
ECU & 2006 & 179 & 82.5\% & 7.4\% & 13.0\% & 5.3\% & 20.8\% \\
ECU & 2010 & 80 & 67.4\% & 10.0\% & 11.6\% & 0.0\% & 0.0\% \\
ECU & 2017 & 92 & 87.1\% & 0.0\% & 0.0\% & 1.0\% & 9.1\% \\
EGY & 2013 & 1039 & 48.0\% & 3.1\% & 6.4\% & 3.4\% & 11.4\% \\
EGY & 2016 & 563 & 50.6\% & 17.1\% & 20.9\% & 14.3\% & 18.9\% \\
EGY & 2020 & 1352 & 67.1\% & 4.8\% & 2.9\% & 3.2\% & 0.6\% \\
SLV & 2006 & 158 & 43.5\% & 5.7\% & 8.1\% & 9.3\% & 17.9\% \\
SLV & 2010 & 67 & 57.3\% & 7.2\% & 4.2\% & 0.0\% & 0.0\% \\
SLV & 2016 & 184 & 49.8\% & 7.3\% & 1.4\% & 6.0\% & 5.8\% \\
SLV & 2023 & 80 & 71.5\% & 0.0\% & 0.0\% & 1.0\% & 0.5\% \\
EST & 2013 & 27 & 50.8\% & 0.0\% & 0.0\% & 0.0\% & 0.0\% \\
EST & 2019 & 71 & 67.5\% & 0.0\% & 0.0\% & 0.0\% & 0.0\% \\
EST & 2023 & 55 & 96.7\% & 0.0\% & 0.0\% & 10.0\% & 0.6\% \\
SWZ & 2006 & 59 & 39.5\% & 2.9\% & 52.2\% & 1.6\% & 43.6\% \\
SWZ & 2016 & 22 & 33.4\% & 30.0\% & 2.6\% & 0.0\% & 0.0\% \\
ETH & 2011 & 35 & 53.3\% & 0.0\% & 0.0\% & 1.0\% & 5.3\% \\
ETH & 2015 & 193 & 61.9\% & 9.2\% & 11.3\% & 5.0\% & 0.6\% \\
FIN & 2020 & 387 & 95.1\% & 0.0\% & 0.0\% & 2.7\% & 0.8\% \\
FRA & 2021 & 394 & 95.2\% & 7.5\% & 3.0\% & 7.0\% & 16.1\% \\
GMB & 2006 & 22 & 22.6\% & 2.7\% & 40.1\% & 18.6\% & 57.6\% \\
GMB & 2018 & 42 & 20.0\% & 15.0\% & 9.5\% & 12.9\% & 30.8\% \\
GMB & 2023 & 52 & 11.1\% & 12.4\% & 6.5\% & 0.0\% & 0.0\% \\
GEO & 2013 & 29 & 28.7\% & 5.0\% & 5.4\% & 0.0\% & 0.0\% \\
GEO & 2019 & 64 & 40.5\% & 0.0\% & 0.0\% & 0.0\% & 0.0\% \\
GEO & 2023 & 94 & 66.7\% & 1.0\% & 0.6\% & 0.0\% & 0.0\% \\
\hline
\hline
\end{tabular}
\end{table}

\begin{table}[]
\begin{tabular}{llllllll}
\hline
\hline
ISO & Year & \begin{tabular}[c]{@{}l@{}}Number \\ of firms\end{tabular} & \begin{tabular}[c]{@{}l@{}}Share of \\ modern \\firms\end{tabular} & \begin{tabular}[c]{@{}l@{}}Bribe in \\ traditional \\ sector\end{tabular} & \begin{tabular}[c]{@{}l@{}}Probability of \\ bribe request in \\ traditional sector\end{tabular} & \begin{tabular}[c]{@{}l@{}}Bribe in \\ modern\\ sector\end{tabular} & \begin{tabular}[c]{@{}l@{}}Probability of \\ bribe request in \\ modern sector\end{tabular} \\
\hline
DEU & 2021 & 337 & 97.4\% & 0.0\% & 0.0\% & 5.1\% & 0.2\% \\
GHA & 2007 & 254 & 11.2\% & 6.3\% & 32.7\% & 6.1\% & 36.3\% \\
GHA & 2013 & 102 & 22.0\% & 7.7\% & 23.8\% & 6.9\% & 17.7\% \\
GHA & 2023 & 67 & 51.8\% & 22.6\% & 23.6\% & 14.8\% & 10.7\% \\
GRC & 2018 & 253 & 90.1\% & 1.0\% & 1.6\% & 2.7\% & 1.4\% \\
GRC & 2023 & 177 & 90.4\% & 1.0\% & 6.5\% & 1.0\% & 0.3\% \\
GRD & 2010 & 17 & 88.6\% & 0.0\% & 0.0\% & 0.0\% & 0.0\% \\
GTM & 2006 & 150 & 55.7\% & 10.1\% & 14.5\% & 41.2\% & 9.3\% \\
GTM & 2010 & 180 & 47.1\% & 23.3\% & 1.4\% & 16.1\% & 6.9\% \\
GTM & 2017 & 66 & 63.9\% & 5.0\% & 1.8\% & 1.0\% & 10.1\% \\
GIN & 2006 & 112 & 6.8\% & 5.9\% & 82.0\% & 9.6\% & 87.9\% \\
GNB & 2006 & 26 & 37.9\% & 7.1\% & 67.2\% & 7.0\% & 69.9\% \\
GUY & 2010 & 45 & 70.1\% & 1.0\% & 5.8\% & 1.2\% & 16.4\% \\
HND & 2006 & 103 & 33.8\% & 10.4\% & 31.6\% & 4.3\% & 18.1\% \\
HND & 2010 & 56 & 65.6\% & 20.0\% & 0.8\% & 18.7\% & 1.2\% \\
HND & 2016 & 41 & 49.9\% & 0.0\% & 0.0\% & 14.8\% & 1.5\% \\
HUN & 2013 & 10 & 35.4\% & 7.2\% & 36.4\% & 0.0\% & 0.0\% \\
HUN & 2019 & 192 & 92.0\% & 20.0\% & 5.6\% & 2.0\% & 5.6\% \\
HUN & 2023 & 190 & 87.9\% & 0.0\% & 0.0\% & 1.0\% & 0.1\% \\
IND & 2014 & 2393 & 42.8\% & 2.4\% & 14.1\% & 4.2\% & 11.4\% \\
IND & 2022 & 3262 & 32.4\% & 6.0\% & 42.0\% & 3.8\% & 36.2\% \\
IDN & 2009 & 429 & 3.9\% & 3.9\% & 7.0\% & 2.1\% & 22.6\% \\
IDN & 2015 & 582 & 13.3\% & 7.6\% & 3.6\% & 16.4\% & 1.6\% \\
IDN & 2023 & 132 & 4.3\% & 14.1\% & 8.0\% & 0.0\% & 0.0\% \\
IRQ & 2011 & 299 & 48.8\% & 3.6\% & 6.9\% & 4.9\% & 19.5\% \\
IRQ & 2022 & 73 & 77.4\% & 50.0\% & 6.6\% & 0.0\% & 0.0\% \\
IRL & 2020 & 132 & 97.6\% & 0.0\% & 0.0\% & 0.0\% & 0.0\% \\
ISR & 2013 & 98 & 83.0\% & 0.0\% & 0.0\% & 0.0\% & 0.0\% \\
ITA & 2019 & 242 & 95.3\% & 5.1\% & 51.2\% & 18.4\% & 1.6\% \\
JAM & 2010 & 63 & 53.7\% & 10.0\% & 4.1\% & 6.9\% & 13.4\% \\
JOR & 2013 & 168 & 51.4\% & 1.6\% & 0.5\% & 1.0\% & 1.0\% \\
JOR & 2019 & 23 & 97.4\% & 2.0\% & 18.8\% & 1.0\% & 3.4\% \\
KAZ & 2009 & 14 & 48.0\% & 8.6\% & 100.0\% & 5.1\% & 100.0\% \\
KAZ & 2013 & 27 & 41.6\% & 2.5\% & 12.8\% & 15.0\% & 13.1\% \\
KAZ & 2019 & 265 & 80.5\% & 6.4\% & 9.8\% & 8.9\% & 2.4\% \\
KEN & 2007 & 361 & 80.2\% & 3.9\% & 66.4\% & 3.1\% & 69.8\% \\
KEN & 2013 & 143 & 76.4\% & 3.5\% & 21.1\% & 5.8\% & 30.6\% \\
KEN & 2018 & 186 & 68.2\% & 8.2\% & 20.7\% & 4.4\% & 22.0\% \\
XKX & 2013 & 22 & 97.3\% & 0.0\% & 0.0\% & 10.0\% & 15.5\% \\
XKX & 2019 & 16 & 93.9\% & 0.0\% & 0.0\% & 2.0\% & 12.5\% \\
KGZ & 2009 & 12 & 34.7\% & 5.3\% & 100.0\% & 7.0\% & 100.0\% \\
KGZ & 2013 & 17 & 31.5\% & 1.3\% & 17.2\% & 10.0\% & 38.0\% \\
KGZ & 2019 & 29 & 60.4\% & 10.0\% & 17.0\% & 10.0\% & 3.9\% \\
KGZ & 2023 & 27 & 49.1\% & 6.7\% & 29.3\% & 8.3\% & 51.1\% \\
\hline
\hline
\end{tabular}
\end{table}

\begin{table}[]
\begin{tabular}{llllllll}
\hline
\hline
ISO & Year & \begin{tabular}[c]{@{}l@{}}Number \\ of firms\end{tabular} & \begin{tabular}[c]{@{}l@{}}Share of \\ modern \\firms\end{tabular} & \begin{tabular}[c]{@{}l@{}}Bribe in \\ traditional \\ sector\end{tabular} & \begin{tabular}[c]{@{}l@{}}Probability of \\ bribe request in \\ traditional sector\end{tabular} & \begin{tabular}[c]{@{}l@{}}Bribe in \\ modern\\ sector\end{tabular} & \begin{tabular}[c]{@{}l@{}}Probability of \\ bribe request in \\ modern sector\end{tabular} \\
\hline
LAO & 2009 & 85 & 19.8\% & 3.1\% & 6.6\% & 1.5\% & 5.0\% \\
LAO & 2016 & 39 & 29.2\% & 2.0\% & 60.6\% & 1.4\% & 44.7\% \\
LAO & 2018 & 20 & 55.4\% & 2.6\% & 28.3\% & 4.7\% & 55.9\% \\
LVA & 2019 & 62 & 53.0\% & 2.0\% & 11.5\% & 0.0\% & 0.0\% \\
LBN & 2013 & 89 & 70.7\% & 1.7\% & 23.3\% & 2.4\% & 32.6\% \\
LBN & 2019 & 172 & 80.4\% & 0.0\% & 0.0\% & 8.6\% & 6.4\% \\
LSO & 2016 & 29 & 30.9\% & 5.6\% & 11.2\% & 0.0\% & 0.0\% \\
LSO & 2023 & 16 & 35.5\% & 0.0\% & 0.0\% & 0.0\% & 0.0\% \\
LBR & 2017 & 43 & 1.6\% & 6.0\% & 28.1\% & 15.0\% & 34.8\% \\
LTU & 2013 & 31 & 61.0\% & 3.0\% & 4.5\% & 1.0\% & 2.3\% \\
LTU & 2019 & 71 & 86.2\% & 0.0\% & 0.0\% & 0.0\% & 0.0\% \\
LUX & 2020 & 19 & 95.7\% & 0.0\% & 0.0\% & 0.0\% & 0.0\% \\
MDG & 2009 & 24 & 29.8\% & 10.1\% & 97.4\% & 13.3\% & 100.0\% \\
MDG & 2022 & 26 & 0.0\% & 6.8\% & 15.6\% & 0.0\% & 0.0\% \\
MWI & 2014 & 26 & 34.7\% & 1.6\% & 25.2\% & 2.0\% & 17.3\% \\
MYS & 2015 & 120 & 48.2\% & 10.6\% & 49.3\% & 24.2\% & 45.5\% \\
MYS & 2019 & 389 & 81.8\% & 11.5\% & 0.6\% & 2.0\% & 0.0\% \\
MLI & 2007 & 258 & 14.8\% & 6.5\% & 21.4\% & 5.4\% & 28.5\% \\
MLI & 2010 & 14 & 3.7\% & 10.0\% & 16.4\% & 2.0\% & 100.0\% \\
MLI & 2016 & 18 & 50.7\% & 10.0\% & 36.1\% & 13.1\% & 92.7\% \\
MLT & 2019 & 44 & 97.5\% & 0.0\% & 0.0\% & 0.0\% & 0.0\% \\
MRT & 2006 & 47 & 79.2\% & 3.2\% & 82.9\% & 8.4\% & 87.8\% \\
MUS & 2023 & 50 & 77.1\% & 0.0\% & 0.0\% & 1.0\% & 0.8\% \\
MEX & 2006 & 509 & 48.4\% & 24.8\% & 9.1\% & 8.1\% & 10.3\% \\
MEX & 2010 & 51 & 49.3\% & 7.0\% & 100.0\% & 2.6\% & 100.0\% \\
MEX & 2023 & 390 & 99.8\% & 0.0\% & 0.0\% & 0.0\% & 0.0\% \\
MDA & 2019 & 43 & 70.1\% & 5.9\% & 21.4\% & 5.0\% & 8.9\% \\
MNG & 2013 & 35 & 36.6\% & 5.0\% & 31.0\% & 8.5\% & 27.3\% \\
MNG & 2019 & 48 & 38.6\% & 2.0\% & 6.1\% & 1.4\% & 11.2\% \\
MNE & 2019 & 27 & 93.6\% & 0.0\% & 0.0\% & 2.7\% & 53.2\% \\
MNE & 2023 & 29 & 95.7\% & 0.0\% & 0.0\% & 1.0\% & 0.4\% \\
MAR & 2013 & 76 & 48.5\% & 1.2\% & 8.5\% & 1.8\% & 9.0\% \\
MAR & 2019 & 161 & 41.1\% & 20.7\% & 15.1\% & 10.0\% & 26.6\% \\
MAR & 2023 & 37 & 65.3\% & 3.0\% & 100.0\% & 5.0\% & 100.0\% \\
MOZ & 2007 & 261 & 30.6\% & 7.0\% & 14.7\% & 7.9\% & 15.8\% \\
MOZ & 2018 & 146 & 25.3\% & 15.1\% & 10.9\% & 4.8\% & 23.7\% \\
MMR & 2014 & 141 & 79.8\% & 11.3\% & 4.6\% & 10.9\% & 9.8\% \\
MMR & 2016 & 257 & 29.7\% & 1.3\% & 11.0\% & 1.2\% & 12.0\% \\
NAM & 2006 & 64 & 80.5\% & 3.0\% & 15.5\% & 7.3\% & 22.3\% \\
NAM & 2014 & 12 & 25.8\% & 3.0\% & 23.2\% & 3.0\% & 33.3\% \\
NPL & 2009 & 74 & 27.2\% & 2.1\% & 5.9\% & 1.1\% & 17.0\% \\
NPL & 2013 & 181 & 34.3\% & 5.3\% & 4.1\% & 6.9\% & 7.3\% \\
NPL & 2023 & 165 & 49.2\% & 10.0\% & 0.5\% & 3.4\% & 2.2\% \\
NLD & 2020 & 271 & 99.2\% & 1.0\% & 19.4\% & 1.7\% & 0.9\% \\
NZL & 2023 & 44 & 99.1\% & 0.0\% & 0.0\% & 0.0\% & 0.0\% \\
\hline
\hline
\end{tabular}
\end{table}

\begin{table}[]
\begin{tabular}{llllllll}
\hline
\hline
ISO & Year & \begin{tabular}[c]{@{}l@{}}Number \\ of firms\end{tabular} & \begin{tabular}[c]{@{}l@{}}Share of \\ modern \\firms\end{tabular} & \begin{tabular}[c]{@{}l@{}}Bribe in \\ traditional \\ sector\end{tabular} & \begin{tabular}[c]{@{}l@{}}Probability of \\ bribe request in \\ traditional sector\end{tabular} & \begin{tabular}[c]{@{}l@{}}Bribe in \\ modern\\ sector\end{tabular} & \begin{tabular}[c]{@{}l@{}}Probability of \\ bribe request in \\ modern sector\end{tabular} \\
\hline
NIC & 2006 & 102 & 16.1\% & 11.5\% & 15.4\% & 8.2\% & 35.1\% \\
NIC & 2010 & 58 & 17.7\% & 2.4\% & 2.0\% & 15.2\% & 4.5\% \\
NIC & 2016 & 62 & 43.5\% & 12.5\% & 7.3\% & 11.0\% & 9.2\% \\
NGA & 2007 & 910 & 14.7\% & 4.5\% & 38.6\% & 4.6\% & 29.7\% \\
NGA & 2014 & 157 & 2.0\% & 10.1\% & 23.4\% & 11.3\% & 42.1\% \\
MKD & 2013 & 59 & 68.2\% & 0.0\% & 0.0\% & 8.1\% & 11.1\% \\
MKD & 2019 & 27 & 54.1\% & 1.0\% & 1.5\% & 0.0\% & 0.0\% \\
MKD & 2023 & 28 & 62.6\% & 10.0\% & 1.0\% & 15.0\% & 16.8\% \\
PAK & 2013 & 149 & 28.7\% & 6.1\% & 12.5\% & 5.0\% & 21.4\% \\
PAK & 2022 & 446 & 70.4\% & 0.0\% & 0.0\% & 22.3\% & 2.0\% \\
PAN & 2006 & 103 & 76.5\% & 6.4\% & 22.8\% & 12.3\% & 23.7\% \\
PNG & 2015 & 11 & 65.1\% & 4.0\% & 100.0\% & 2.1\% & 100.0\% \\
PRY & 2006 & 43 & 61.4\% & 4.8\% & 79.6\% & 14.1\% & 60.7\% \\
PRY & 2010 & 41 & 70.0\% & 10.0\% & 9.2\% & 1.5\% & 13.4\% \\
PRY & 2017 & 29 & 96.4\% & 0.0\% & 0.0\% & 17.0\% & 20.0\% \\
PRY & 2023 & 37 & 80.6\% & 0.0\% & 0.0\% & 1.0\% & 7.7\% \\
PER & 2006 & 137 & 46.5\% & 4.1\% & 4.2\% & 3.9\% & 5.2\% \\
PER & 2010 & 425 & 61.9\% & 5.7\% & 20.9\% & 9.0\% & 21.1\% \\
PER & 2017 & 224 & 73.8\% & 11.5\% & 11.7\% & 9.2\% & 18.7\% \\
PER & 2023 & 221 & 66.3\% & 5.8\% & 2.3\% & 5.1\% & 8.3\% \\
PHL & 2009 & 276 & 43.0\% & 7.4\% & 20.1\% & 4.9\% & 25.5\% \\
PHL & 2015 & 144 & 34.2\% & 2.4\% & 37.6\% & 2.4\% & 35.8\% \\
PHL & 2023 & 180 & 32.5\% & 18.0\% & 0.5\% & 29.1\% & 1.8\% \\
POL & 2013 & 15 & 83.1\% & 0.0\% & 0.0\% & 0.0\% & 0.0\% \\
POL & 2019 & 128 & 91.5\% & 25.0\% & 34.6\% & 0.0\% & 0.0\% \\
PRT & 2019 & 448 & 68.9\% & 50.0\% & 0.3\% & 0.0\% & 0.0\% \\
PRT & 2023 & 232 & 96.3\% & 0.0\% & 0.0\% & 2.0\% & 0.2\% \\
ROU & 2013 & 64 & 62.3\% & 2.4\% & 15.5\% & 1.0\% & 4.3\% \\
ROU & 2019 & 338 & 38.5\% & 7.9\% & 8.9\% & 9.1\% & 7.6\% \\
ROU & 2023 & 301 & 74.0\% & 0.0\% & 0.0\% & 4.9\% & 2.3\% \\
RUS & 2009 & 44 & 56.7\% & 3.6\% & 99.1\% & 5.0\% & 100.0\% \\
RUS & 2012 & 260 & 56.1\% & 4.7\% & 14.8\% & 12.5\% & 8.8\% \\
RUS & 2019 & 377 & 76.8\% & 10.2\% & 1.9\% & 8.2\% & 2.2\% \\
RWA & 2006 & 48 & 45.5\% & 8.9\% & 18.7\% & 4.3\% & 15.4\% \\
RWA & 2019 & 83 & 38.5\% & 0.0\% & 0.0\% & 1.0\% & 1.3\% \\
RWA & 2023 & 68 & 32.5\% & 5.3\% & 23.7\% & 2.0\% & 1.6\% \\
WSM & 2023 & 29 & 55.1\% & 0.0\% & 0.0\% & 0.0\% & 0.0\% \\
SAU & 2022 & 454 & 40.0\% & 0.0\% & 0.0\% & 0.0\% & 0.0\% \\
SEN & 2007 & 238 & 39.3\% & 6.0\% & 28.8\% & 9.6\% & 22.3\% \\
SEN & 2014 & 58 & 53.0\% & 8.9\% & 12.8\% & 4.0\% & 4.2\% \\
SRB & 2009 & 11 & 67.7\% & 9.8\% & 100.0\% & 6.6\% & 100.0\% \\
SRB & 2013 & 30 & 62.2\% & 5.0\% & 40.0\% & 2.7\% & 32.8\% \\
SRB & 2019 & 26 & 90.2\% & 0.0\% & 0.0\% & 5.0\% & 7.6\% \\
SLE & 2017 & 38 & 100.0\% & 0.0\% & 0.0\% & 21.5\% & 43.8\% \\
SLE & 2023 & 71 & 0.0\% & 27.3\% & 14.7\% & 0.0\% & 0.0\% \\
\hline
\hline
\end{tabular}
\end{table}

\begin{table}[]
\begin{tabular}{llllllll}
\hline
\hline
ISO & Year & \begin{tabular}[c]{@{}l@{}}Number \\ of firms\end{tabular} & \begin{tabular}[c]{@{}l@{}}Share of \\ modern \\firms\end{tabular} & \begin{tabular}[c]{@{}l@{}}Bribe in \\ traditional \\ sector\end{tabular} & \begin{tabular}[c]{@{}l@{}}Probability of \\ bribe request in \\ traditional sector\end{tabular} & \begin{tabular}[c]{@{}l@{}}Bribe in \\ modern\\ sector\end{tabular} & \begin{tabular}[c]{@{}l@{}}Probability of \\ bribe request in \\ modern sector\end{tabular} \\
\hline
SGP & 2023 & 94 & 45.4\% & 0.0\% & 0.0\% & 0.0\% & 0.0\% \\
SVK & 2013 & 17 & 93.1\% & 0.0\% & 0.0\% & 7.7\% & 13.9\% \\
SVK & 2019 & 125 & 83.2\% & 7.7\% & 26.5\% & 6.9\% & 22.7\% \\
SVK & 2023 & 82 & 93.9\% & 0.0\% & 0.0\% & 10.0\% & 3.3\% \\
SVN & 2013 & 45 & 90.7\% & 7.9\% & 31.6\% & 10.3\% & 25.1\% \\
SVN & 2019 & 51 & 98.1\% & 0.0\% & 0.0\% & 5.0\% & 4.1\% \\
SLB & 2015 & 13 & 60.7\% & 1.3\% & 79.5\% & 1.0\% & 73.0\% \\
ZAF & 2007 & 611 & 75.2\% & 6.8\% & 11.6\% & 4.3\% & 9.1\% \\
ZAF & 2020 & 231 & 79.5\% & 3.0\% & 1.3\% & 1.0\% & 0.1\% \\
SSD & 2014 & 31 & 50.0\% & 12.5\% & 11.9\% & 9.0\% & 23.9\% \\
ESP & 2021 & 506 & 99.8\% & 0.0\% & 0.0\% & 0.0\% & 0.0\% \\
LKA & 2011 & 195 & 34.4\% & 9.1\% & 2.1\% & 1.2\% & 1.4\% \\
KNA & 2010 & 17 & 76.1\% & 0.0\% & 0.0\% & 0.0\% & 0.0\% \\
LCA & 2010 & 42 & 100.0\% & 0.0\% & 0.0\% & 0.0\% & 0.0\% \\
VCT & 2010 & 40 & 97.0\% & 0.0\% & 0.0\% & 1.5\% & 5.1\% \\
SUR & 2010 & 71 & 91.4\% & 0.0\% & 0.0\% & 5.9\% & 14.1\% \\
SUR & 2018 & 20 & 51.5\% & 5.2\% & 55.2\% & 4.0\% & 38.0\% \\
SWE & 2020 & 260 & 99.1\% & 0.0\% & 0.0\% & 0.0\% & 0.0\% \\
TJK & 2008 & 10 & 32.6\% & 10.7\% & 100.0\% & 11.5\% & 100.0\% \\
TJK & 2013 & 16 & 26.3\% & 3.4\% & 39.0\% & 10.0\% & 66.5\% \\
TZA & 2006 & 222 & 47.5\% & 6.9\% & 39.8\% & 6.4\% & 53.8\% \\
TZA & 2013 & 98 & 40.6\% & 1.5\% & 3.0\% & 0.0\% & 0.0\% \\
TZA & 2023 & 111 & 3.6\% & 6.5\% & 4.6\% & 28.6\% & 27.9\% \\
THA & 2016 & 395 & 20.9\% & 13.4\% & 0.9\% & 0.0\% & 0.0\% \\
TLS & 2015 & 29 & 11.2\% & 2.3\% & 85.8\% & 2.8\% & 100.0\% \\
TGO & 2016 & 15 & 62.8\% & 1.0\% & 16.7\% & 1.0\% & 8.9\% \\
TGO & 2023 & 33 & 73.7\% & 0.0\% & 0.0\% & 0.0\% & 0.0\% \\
TTO & 2010 & 70 & 93.8\% & 1.0\% & 11.2\% & 2.4\% & 10.5\% \\
TUN & 2013 & 151 & 84.1\% & 0.0\% & 0.0\% & 2.0\% & 0.3\% \\
TUN & 2020 & 74 & 55.6\% & 0.0\% & 0.0\% & 0.0\% & 0.0\% \\
UGA & 2006 & 232 & 35.7\% & 7.2\% & 42.9\% & 6.5\% & 55.4\% \\
UGA & 2013 & 31 & 20.5\% & 14.0\% & 9.0\% & 5.0\% & 26.0\% \\
UKR & 2008 & 30 & 73.4\% & 7.4\% & 100.0\% & 6.2\% & 100.0\% \\
UKR & 2013 & 42 & 47.5\% & 5.5\% & 95.1\% & 8.1\% & 95.5\% \\
UKR & 2019 & 182 & 60.0\% & 3.7\% & 13.7\% & 4.6\% & 6.6\% \\
URY & 2006 & 23 & 68.9\% & 3.0\% & 6.7\% & 8.3\% & 6.8\% \\
URY & 2010 & 129 & 63.6\% & 2.0\% & 3.8\% & 2.2\% & 2.7\% \\
URY & 2017 & 34 & 60.4\% & 0.0\% & 0.0\% & 0.0\% & 0.0\% \\
UZB & 2008 & 34 & 23.1\% & 3.9\% & 100.0\% & 3.8\% & 100.0\% \\
UZB & 2013 & 47 & 59.7\% & 2.9\% & 9.0\% & 4.7\% & 4.6\% \\
UZB & 2019 & 215 & 62.3\% & 10.7\% & 4.6\% & 7.3\% & 8.2\% \\
VEN & 2010 & 24 & 93.7\% & 8.6\% & 83.7\% & 9.6\% & 15.2\% \\
PSE & 2013 & 105 & 64.7\% & 16.7\% & 7.2\% & 24.1\% & 6.8\% \\
PSE & 2019 & 66 & 72.1\% & 3.0\% & 0.2\% & 14.8\% & 3.7\% \\
PSE & 2023 & 102 & 60.6\% & 18.6\% & 23.2\% & 7.4\% & 12.1\% \\
\hline
\hline
\end{tabular}
\end{table}

\begin{table}[]
\begin{tabular}{llllllll}
\hline
\hline
ISO & Year & \begin{tabular}[c]{@{}l@{}}Number \\ of firms\end{tabular} & \begin{tabular}[c]{@{}l@{}}Share of \\ modern \\firms\end{tabular} & \begin{tabular}[c]{@{}l@{}}Bribe in \\ traditional \\ sector\end{tabular} & \begin{tabular}[c]{@{}l@{}}Probability of \\ bribe request in \\ traditional sector\end{tabular} & \begin{tabular}[c]{@{}l@{}}Bribe in \\ modern\\ sector\end{tabular} & \begin{tabular}[c]{@{}l@{}}Probability of \\ bribe request in \\ modern sector\end{tabular} \\
\hline
YEM & 2010 & 53 & 15.8\% & 7.1\% & 61.7\% & 11.6\% & 34.7\% \\
YEM & 2013 & 37 & 40.6\% & 1.9\% & 59.6\% & 2.7\% & 22.7\% \\
ZMB & 2007 & 272 & 51.6\% & 7.4\% & 12.9\% & 1.9\% & 11.8\% \\
ZMB & 2013 & 122 & 33.6\% & 10.1\% & 7.3\% & 6.1\% & 13.7\% \\
ZMB & 2019 & 13 & 58.1\% & 11.7\% & 80.7\% & 5.0\% & 43.4\% \\
ZWE & 2016 & 100 & 57.4\% & 8.4\% & 5.4\% & 7.1\% & 8.5\% \\
\hline
\hline
\end{tabular}
\end{table}
 
\end{document}